\documentclass[prd,superscriptaddress,showpacs]{revtex4}
\usepackage{amsmath}
\usepackage{graphicx}
\def \D {\tilde{\nabla}}
\def \curl {\mbox{curl}\,}
\def \ep {\varepsilon}
\def\l{\label}

\def\Th{\Theta}
\def\dd{\mathcal D}
\def\zz{\mathcal Z}
\def\sig{\sigma}
\def\om{\omega}
\def\udot{\dot{u}}
\def\nab{\nabla}
\def\3nab{\tilde{\nabla}}

\def\lgl{\langle}
\def\rgl{\rangle}

\def\nn{\nonumber}
\def\c{\mbox{curl}}

\def\hsp5{\hspace{5mm}}

\def\case#1/#2{\textstyle\frac{#1}{#2}}

\def\be {\begin{equation}}
\def\ee {\end{equation}}
\def\ber {\begin{eqnarray}}
\def\eer {\end{eqnarray}}
\def\bea {\begin{eqnarray}}
\def\eea {\end{eqnarray}}

\def\bc {\begin{center}}
\def\ec {\end{center}}
\def\case#1/#2{\frac{#1}{#2}}
\def\rf#1{(\ref{#1})}
\newcommand{\n}{{}^{(3)}\nabla} \newcommand{\na}{\nabla}

\newcommand{\hs}{\,-\,}

\newcommand{\0}{^{(0)}}
\newcommand{\1}{^{(1)}}
\newcommand{\2}{^{(2)}}
\newcommand{\ab}{_{\alpha\beta}}

\def\cqg{{\it Class. Quantum Grav.}\ }

\def\grg{{\it Gen. Relativ. Grav.}\ }

\def\etal\;{{\it et al.}}

\begin{document}

\title{The evolution of density perturbations in $f(R)$ gravity}

\author{S. Carloni\footnote{\tt
sante.carloni@gmail.com}}
\affiliation{ Department of Mathematics
and Applied\ Mathematics, University of Cape Town, South Africa.}
\author{P. K. S. Dunsby\footnote{\tt peter.dunsby@uct.ac.za}}
\affiliation{ Department of Mathematics and Applied\ Mathematics,
University of Cape Town, South Africa.}
\affiliation{\ South African
Astronomical Observatory, Observatory Cape Town, South Africa.}
\author{A. Troisi\footnote{\tt antrois@gmail.com}}
\affiliation{ Dipartimento di Scienze Fisiche e Sez. INFN di Napoli,
Universita' di Napoli ``Federico II", Complesso Universitario di
Monte S. Angelo, Via Cinthia I-80126 Napoli (Italy)}

\begin{abstract}
We give a rigorous and mathematically well defined presentation of
the Covariant and Gauge Invariant theory of scalar perturbations of
a Friedmann-Lema\^{\i}tre-Robertson-Walker universe for Fourth Order
Gravity, where the matter is described by a perfect fluid with a
barotropic equation of state. The general perturbations equations
are applied to a simple background solution of $R^n$ gravity. We
obtain exact solutions of the perturbations equations for scales
much bigger than the Hubble radius. These solutions have a number of
interesting features. In particular, we find that for all values of
n there is always a growing mode for the density contrast, even if
the universe undergoes an accelerated expansion. Such a behaviour
does not occur in standard General Relativity, where as soon as Dark
Energy dominates, the density contrast experiences an unrelenting
decay. This peculiarity is sufficiently novel to warrant further
investigation of fourth order gravity models.
\end{abstract}

\date{\today}
\pacs{04.50.+h, 04.25.Nx } \maketitle

\section{Introduction}
In the last few years the idea of a geometrical origin for Dark Energy (DE) i.e.
the connection between DE and a non-standard behavior of gravitation on
cosmological scales has attracted a considerable amount of interest.

Higher order gravity, and in particular fourth order gravity, has
been widely studied in the case of the
Friedmann-Lema\^{\i}tre-Robertson-Walker (FLRW) metric using a
number of different techniques (see for example \cite{revnostra,
Odintsov,Carroll,Capozziello:2005ku,Capozziello:2006dj,Capozziello:2006ph,Capozziello:2006jj}).
Recently a general approach was developed to analyze the phase space
of the fourth order cosmologies
\cite{cdct:dynsys05,SanteGenDynSys,shosho}, providing for the first
time a way of obtaining exact solutions together with their
stability and a general idea of the qualitative behavior of these
cosmological models.

The phase space analysis shows that for  FLRW models there exist
classes of fourth order theories which admit a  transient
decelerated expansion phase, followed by one with an accelerated
expansion rate (see also \cite{Capozziello:2006dj} for a different
approach). The first (Friedmann-like) phase provides a setting
during which structure formation can take place, followed by a
smooth transition to a DE-like era which drives the cosmological
acceleration. However, in order to determine if this is the case, we
need to develop a theory of cosmological perturbations for higher
order gravity.

The aim of this paper is to give a rigorous and mathematically well
defined theory which describes the evolution scalar perturbations of FLRW
models in fourth order gravity, which can be used to investigate this issue in detail
(see \cite{current1,current2, current3,current4} for other recent contributions to this
area).

In order to achieve this goal a perturbation formalism needs to be chosen that is
best suited for this task. One possible choice is the Bardeen metric based approach \cite{bi:bardeen}
which guarantees the gauge invariance of the results. However this approach has the
drawback of introducing variables which only have a clear physical meaning in certain gauges
\cite{BDE}.  Although this is not a big problem in the context of General Relativity (GR), this is not
necessarily true in  the case of higher order gravity and consequently could lead to a
miss-interpretation of the results.

In what follows we will use, instead, the covariant and gauge invariant
approach developed for GR in \cite{EB,EBH,BDE,DBE,BED,DBBE} which has the
advantage of using perturbation variables with a clear geometrical and physical
interpretation. Furthermore, we use a specific recasting of the field
equations that will make the development of the cosmological perturbation
theory even more transparent.

The outline of the paper is as follows. In Section II we give the general
set up of the equations. In Section III we introduce the 1+3
covariant approach and apply it to the field equations presented
in Section I. In Section IV we give the general evolution equations for
the 1+3 quantities and linearize them around a FLRW spacetime. In
Section V we introduce the perturbation variables and their
propagation equations giving also their harmonic decomposition. In
Section VI we apply these equation to the case of $R^{n}$-gravity with a
barotropic perfect fluid matter source and find exact solution in long wavelength limit.
Finally in Section VII we present our conclusions.

Unless otherwise specified, natural units ($\hbar=c=k_{B}=8\pi G=1$)
will be used throughout this paper, Latin indices run from 0 to 3.
The symbol $\nabla$ represents the usual covariant derivative and
$\partial$ corresponds to partial differentiation. We use the
$-,+,+,+$ signature and the Riemann tensor is defined by
\begin{equation}
R^{a}{}_{bcd}=W^a{}_{bd,c}-W^a{}_{bc,d}+ W^e{}_{bd}W^a{}_{ce}-
W^f{}_{bc}W^a{}_{df}\;,
\end{equation}
where the $W^a{}_{bd}$ are the Christoffel symbols (i.e. symmetric in
the lower indices), defined by
\begin{equation}
W^a_{bd}=\frac{1}{2}g^{ae}
\left(g_{be,d}+g_{ed,b}-g_{bd,e}\right)\;.
\end{equation}
The Ricci tensor is obtained by contracting the {\em first} and the
{\em third} indices
\begin{equation}\label{Ricci}
R_{ab}=g^{cd}R_{acbd}\;.
\end{equation}
Finally the Hilbert--Einstein action in the presence of matter is
given by
\begin{equation}
{\cal A}=\int d x^{4} \sqrt{-g}\left[\frac{1}{2}R+ L_{m}\right]\;.
\end{equation}

\section{General equations for fourth order gravity.}
The  most general action for a fourth order theory of gravity is
given by
\begin{equation}\label{azione generale high order}
\mathcal{A} = \int d^4 x \sqrt{-g} \left[ \Lambda + c_{0} R + c_{1}
R^{2} + c_{2} R_{\mu \nu} R^{\mu \nu} + {\cal L}_{m} \right]\;,
\end{equation}
where we have used the Gauss Bonnet theorem \cite{GB}  and
$\mathcal{L}_m$ represents the matter contribution. In situations where the
metric has a high degree of symmetry, this action can be further simplified. In
particular, in the homogeneous and isotropic case the \rf{azione
generale high order} can be expressed as
\begin{equation}\label{lagr f(R)}
\mathcal{A}=\int d^4 x \sqrt{-g}\left[ f(R)+{\cal L}_{m}\right]\;.
\end{equation}
Varying the action with respect to the metric gives the generalization
of the Einstein equations:
\begin{equation}\label{eq:einstScTn}
f'G_{ab}=f'\left(R_{ab}-\frac{1}{2}\,g_{ab} R\right)=T
_{ab}^{m}+\frac{1}{2}g_{ab} \left(R-R f'\right) +\nab_b\nab_a f'-
g_{ab}\nab_c\nab^c f'\;,
\end{equation}
where $f=f(R)$, $f'= \displaystyle{\frac{d f(R)}{dr}}$, and
$\displaystyle{T^{M}_{\mu\nu}=\frac{2}{\sqrt{-g}}\frac{\delta
(\sqrt{-g}\mathcal{L}_{m})}{\delta g_{\mu\nu}}}$ represents the
stress energy tensor of standard matter. These equations reduce to
the standard Einstein field equations when $f(R)=R$. It is crucial
for our purposes to be able to write \rf{eq:einstScTn} in the form
\begin{equation}
\label{eq:einstScTneff}
 G_{ab}=\tilde{T}_{ab}^{m}+T^{R}_{ab}=T^{tot}_{ab}\,,
 \end{equation}
where $\displaystyle{\tilde{T}_{ab}^{m}=\frac{ T_{ab}^{m}}{f'}}$ and
\begin{eqnarray}\label{eq:TenergymomentuEff}
T_{ab}^{R}=\frac{1}{f'}\left[\frac{1}{2}g_{ab} \left(R-R f'\right)
+\nab_b\nab_a f- g_{ab}\nab_c\nab^cf\right], \label{eq:semt}
\end{eqnarray}
represent two effective ``fluids": the  {\em curvature ``fluid"}
(associated with $T^{R}_{ab}$) and  the {\em effective matter
``fluid"} (associated with $\tilde{T}_{ab}^{m}$). This step is
important because it allows us to treat fourth order gravity as
standard Einstein gravity in the presence of two ``effective"
fluids. This means that once the effective thermodynamics of these
fluids has been studied, we can apply the covariant gauge invariant
approach in the standard way.

The conservation properties of these effective fluids are given by
the Bianchi identities $T_{ab}^{tot\; ;b}$. When applied to the
total stress energy tensor, these identities reveal that if standard
matter is conserved, the total fluid is also conserved even though
the curvature fluid may in general possess off--diagonal terms
\cite{cdct:dynsys05,Taylor,eddington book}. In other words, no
matter how complicated the effective stress energy tensor
$T^{tot}_{ab}$ is, it will always be divergence free if
$T_{ab}^{m;b}=0$. When applied to the single effective
tensors, the Bianchi identities read
\begin{eqnarray}\label{Bianchi}
  &&\tilde{T}_{a b}^{M;b}=\frac{T_{ab}^{m;b}}{f'}-\frac{f''}{f'^{2}}\;T^{m}_{ab}\;R^{; b}\;,\\
  && T_{ab}^{R;b}=\frac{f''}{f'^{2}}\;\tilde{T}^{M}_{ab}\;R^{; b}\;,
\end{eqnarray}
with the last expression being a consequence of total energy-momentum
conservation. It follows that the individual effective fluids are
not conserved but exchange energy and momentum.

It is worth noting here that even if the energy-momentum tensor
associated with the effective matter source is not conserved, standard matter still
follows the usual conservation equations $T_{ab}^{m\ ;b}=0$. It is
also important to stress that the fluids with $T^R_{ab}$ and $\tilde{T}^m_{ab}$
defined above are {\em effective} and consequently can admit  features that one would
normally consider un-physical for a standard matter field. This
means that all the thermodynamical quantities associated with
the curvature defined below should be considered {\em effective}
and not bounded by the usual constraints associated with matter fields.
It is important to understand that this does not compromise any of the
thermodynamical properties  of standard matter represented by the
Lagrangian ${\cal L}_{m}$.

\section{Covariant decomposition of higher order gravity}
In this section we will describe the general covariant decomposition
fourth order gravity. This procedure will take place in two steps.
The first one is to develop the kinematics of the spacetime and  the
second one is to study the thermodynamics of the effective fluids
defined in the previous section.
\subsection{Preliminaries}
The starting point for any  analysis using the covariant approach
is the choice of a suitable frame i.e. the 4-velocity $u_{a}$ of an
observer in spacetime.

This choice changes the structure of the equations and can
simplify the calculations in the same way a choice of coordinates
makes life easier in classical mechanics. Even if the covariance of
the theory guarantees that all velocity fields are  equivalent, a number of natural
choices for $u_{a}$ exist; they are the {\em energy frame} $u^{E}_a$ , which is
defined to be a timelike eigenvector of the stress energy tensor $T_{ab}$, the
{\em particle frame} $u^{N}_a$  that is derived from the particle flux
vector $N_{a}$ and the {\em entropy frame} $u^{S}_a$  defined by the
entropy flux vector  $S_a$.  These frames have the advantage of
inducing important simplifications to the equations (e.g. in the
energy frame the  total energy flux is always zero \cite{BDE}).

The vectors $u^{E}_a, u^{N}_a, u^{S}_a$ always exist in the case of
a perfect fluid and coincide, defining a unique hydrodynamical 4-velocity
for the flow  \cite{BDE}. In the case of more than one perfect
fluid, one can in principle define these three frames for each
component as well as for the total fluid and then choose the most
convenient frame to work in.

In our specific case, equation \rf{eq:einstScTneff} allows us to
define two ``effective" fluids, but their structure does not
necessarily make any of the three frames defined above a suitable
choice. This because both the curvature fluid and the effective
matter do not necessarily satisfy the Weak Energy Condition  (WEC)
$T_{ab}V^{a}V^{b}\geq 0$, $V^{a}V_{a}\leq 0$ \cite{HawkEllis}. This
relation, which for  homogeneous and isotropic spacetimes
corresponds to the requirement that the energy density is positive,
is the key hypothesis which allows the timelike vectors $u^{E}_a,
u^{N}_a, u^{S}_a$ to exist and is, in general, a very reasonable
assumption. In our case, however, the violation of this condition
means that none of the energy entropy or particle frames are, in
general suitable choices of frame.

However, an alternative frame choice follows from the fact that
whatever the behavior of the effective fluid is,  standard matter
is still thermodynamically well defined and consequently the
stress energy tensor $\tilde{T}_{ab}$ satisfies the standard
energy conditions. It  follows that a natural choice of frame
the one of those associated with  standard matter ($u_a= u^{m}_a$),
assumed to be a barotropic perfect fluid with equation of state $p=w\mu$.
\footnote{Another possible choice of frame would be
$u^{R}_a=\nabla_a R$ in analogy with the scalar filed case
\cite{DBBE,ScSante}. However, this choice would imply that in the
3-spaces orthogonal to $u^{R}_a$, the projected spatial gradients
of $R$ vanish, putting a further condition on the structure of the
perturbed spacetime which may not satisfied in general.}.
This choice is also motivated by the fact that the real observers
are attached to galaxies and these galaxies follow the standard
matter geodesics. Consequently this frame choice is the one
which can be best motivated  from a physical point of view.

\subsection{Kinematics}
Once the frame has been chosen the derivation of the kinematical
quantities can be obtained in a standard way \cite{EllisCovariant}. The
derivative along the matter fluid flow lines is defined by $\dot{X}=u_a\nabla^aX$.
The projection tensor into the tangent 3-spaces orthogonal to the
flow vector is:
\begin{equation}
h_{ab}  \equiv g_{ab}+u_au_b\; \Rightarrow h^a{}_b
h^b{}_c=h^a{}_c\;, ~h_{ab}u^b=0\; ,
\end{equation}
and the tensor $\nab_b u_a$ can be expanded as
\begin{equation}
\nab_b u_a=\3nab_b u_a-a_a u_b\;, ~~~\3nab_b u_a=\frac{1}{3}\Theta
h_{ab} +\sigma_{ab}+\omega_{ab}\;,  \label{eq:dec}
\end{equation}
where $\3nab_a=h^b{}_a\nabla_b$ is the spatially totally projected covariant
derivative operator orthogonal to $u^a$. This relation allows us to
define the key kinematic quantities of the cosmological model: the
expansion $\Theta$, the shear $\sigma_{ab}$, the vorticity
$\omega_{ab}$ and the acceleration  $a_a = \dot{u}_a$.

In the following, angle brackets applied to a vector  denote
the projection of this vector on the tangent 3-spaces
\begin{equation}
V_{\langle{{a}}\rangle}=h_{{a}}{}^{{b}} V_{{b}}\;.
\end{equation}
Instead when applied to a tensor they denote the projected,
symmetric and trace free part of this object
\begin{equation}
W_{\langle{{a}}{{b}}\rangle}=\left[h_{({{a}}}{}^{c}
h_{{{b}})}{}^{d}-
{\textstyle\frac{1}{3}}h^{{c}{d}}h_{{{a}}{{b}}}\right]W_{{c}{d}}\,.
\end{equation}
Finally the spatial curl of a variable is
\begin{equation}
(\c\,X)^{ab} = \eta^{cd\lgl a}\,\3nab_{c}X^{b\rgl}\!_{d}
\end{equation}
where $\epsilon_{abc}=u^d\eta_{abcd}$ is the spatial volume.

The general propagation equations for these kinematic variables, for any spacetime
corresponds to the so called {\em 1+3 covariant equations} \cite{EllisCovariant}
and are gven in Appendix \ref{CovID}.
\subsection{Effective total energy-momentum tensors}
The choice of the frame also allows us  to obtain an irreducible
decomposition of the stress energy momentum tensor. In a general
frame and for a general tensor $T_{{a}{{b}}}$ one obtains:
\begin{equation}\label{Tdecomp}
T_{{a}{{b}}}=\mu u_a
u_{{b}}+ph_{{a}{{b}}}+2q_{(a}u_{{{b}})}+\pi_{{{a}}{{b}}}\,,
\end{equation}
where $\mu$ and $p$ are the energy density and isotropic pressure,
$q_{{{a}}}$ is the energy flux ($q_{{{a}}}=q_{\langle{{a}}\rangle}$)  and $\pi_{{{a}}{{b}}}$
is the anisotropic pressure ($\pi_{{{a}}{{b}}}=\pi_{\langle{{a}}{{b}}\rangle}$).

This decomposition can be applied to our effective energy momentum
tensors. Relative to $u_a^m$ we obtain
\begin{eqnarray}\label{mutot}
\mu^{\rm tot}\,&=&T^{\rm tot}_{ab}u^{a}u^{b}\,=\,\tilde{\mu}^{\,
m}+\mu^{\,R}\,,\qquad
p^{\rm tot}\,=\frac{1}{3}T^{\rm tot}_{ab}h^{ab}\,=\,\tilde{p}^{\, m}+p^{\,R}\;,\\
q^{\rm tot}_{a}\,&=&-T^{\rm
tot}_{bc}h_{a}^{b}u^{c}\,=\,\tilde{q}^{\,
m}_{a}+q^{\,R}_{a}\,,\qquad \pi^{\rm tot}_{ab}\,=\,T^{\rm
tot}_{cd}h_{<a}^{c}h_{b>}^{d}\,=\,\tilde{\pi}^{\,
m}_{ab}+\pi^{\,R}_{ab}\,,
\end{eqnarray}
with
\begin{eqnarray}
\tilde{\mu}^{\,m}\,&=&\,\frac{\mu^{\,m}}{f'}\,,\qquad
\tilde{p}^{\,m}\,=\,\frac{p^{\,m}}{f'}\,,\qquad
\tilde{q}^{\,m}_{a}\,=\,\frac{q^{\,m}_{a}}{f'}\,,\qquad
\tilde{\pi}^{\,m}_{ab}=\,\frac{\pi^{\,m}_{ab}}{f'}\, .
\end{eqnarray}
Since we assume that  standard matter is a perfect fluid,
$q^{\,m}_{a}$ and $\pi^{\,m}_{ab}$ are zero, so that the last two
quantities above  also vanish.

The effective thermodynamical quantities for the curvature ``fluid"
are
\begin{eqnarray}
&&\mu^{R}\,=\,\frac{1}{f'}\left[\frac{1}{2}(R f'-f)-\Theta
f''\dot{R}+f''\tilde{\nabla}^2{R}+f''\,\dot{u}_b\D{R}\right]\;,\\
&&p^{R}\,=\,\frac{1}{f'}\left[\frac{1}{2}(f-R
f')+f''\ddot{R}+3f'''\dot{R}^2+\frac{2}{3}\Theta
f''\dot{R}-\frac{2}{3}f''\tilde{\nabla}^2{R} +\right. \nonumber\\
&& \qquad\left.  -\frac{2}{3}f'''\D^{a}{R}\D_{a}{R}-\frac{1}{3}f''
\,\dot{u}_b\D{R}\right]\;,\\
&&q^{R}_a\,=\,-\frac{1}{f'}\left[f'''\dot{R}\D_{a}R+f''\D_{a}\dot{R}-\frac{1}{3}f''
\D_{a}R\right]\;,\\
&&\pi^{R}_{ab}\,=\,\frac{1}{f'}\left[f''\D_{\lgl
a}\D_{b\rgl}R+f'''\D_{\lgl a}{R}\D_{b\rgl}{R}+\sigma_{a
b}\dot{R}\right]\,.\label{piR}
\end{eqnarray}
The twice contracted Bianchi Identities lead to evolution equations for
$\mu^{\,m}$, $\mu^{R}$, $q^{R}_a$:

\begin{eqnarray}
&&\dot{\mu}^m\,=\, - \,\Theta\,(\mu^m+{p^m})\;,\label{eq:cons1}\\
&&\dot{\mu}^R + \3nab^{a}q^R_{a} = - \,\Th\,(\mu^R+p^R)
- 2\,(\udot^{a}q^R_{a}) -
(\sig^{a}\!^{b}\pi^R_{b}\!_{a})+\mu^{m}\frac{f''\,\dot{R}}{f'^{2}}\;,\label{eq:cons2} \\
&& \dot{q}^R_{\lgl a\rgl} + \3nab_{a}p^R +
\3nab^{b}\pi^R_{ab} = - \,{\textstyle\frac{4}{3}}\,\Th\,q^R_{a} -
\sig_{a}\!^{b}\,q^R_{b} - (\mu^R+p^R)\,\udot_{a} -
\udot^{b}\,\pi^R_{ab} -
\eta_{a}^{bc}\,\om_{b}\,q^R_{c}+\mu^{m}\frac{f''\,\D_{a}{R}}{f'^{2}}\l{eq:cons3} 
\ , \end{eqnarray}
and a relation connecting the acceleration $\dot{u}_{a}$ to $\mu^m$
and $p^m$ follows from momentum conservation of standard
matter:
\begin{equation}\label{eq:cons4}
\3nab^{a}{p^m} =  - (\mu^m+{p^m})\,\udot^{a}\,.
\end{equation}
Note that, as we have seen in the previous section the curvature
fluid and the effective matter  exchange energy and momentum. The
decomposed interaction terms in Equations \rf{eq:cons2} and
\rf{eq:cons1} are given by $\mu^{m}\frac{f''\,\D_{a}{R}}{f'^{2}}$
and $\mu^{m}\frac{f''\,\dot{R}}{f'^{2}}$.
\section{Propagation and constraint equations}
\subsection{Nonlinear equations}
We are now ready to write the full nonlinear 1+3 equations for
higher order gravity. Substituting the quantities given above into
the equations given in the Appendix A, we
obtain the following results:\newline

\noindent Expansion propagation (generalized Raychaudhuri equation):
\begin{eqnarray}\label{1+3eqRayHO}
&&\dot{\Theta}+{\textstyle\frac{1}{3}}\Theta^2+\sigma_{{{a}}{{b}}}
\sigma^{{{a}}{{b}}} -2\omega_{{a}}\omega^{{a}} -\tilde{\nabla}^a
\dot{u}_{{a}}+ \dot{u}_{{a}}
\dot{u}^{{a}}+{\textstyle\frac{1}{2}}(\tilde{\mu}^{m} +
3\tilde{p}^{m}) =-{\textstyle\frac{1}{2}}({\mu}^{R} +
3{p}^{R})\;,
\end{eqnarray}
Vorticity propagation:
\begin{equation}\label{1+3VorHO}
\dot{\omega}_{\langle {{a}}\rangle }
+{\textstyle\frac{2}{3}}\Theta\omega_{{a}} +{\textstyle\frac{1}{2}}\curl
\dot{u}_{{a}} -\sigma_{{{a}}{{b}}}\omega^{{b}}=0 \;,
\end{equation}
Shear propagation:
\begin{equation}\label{1+3ShearHO}
\dot{\sigma}_{\langle {{a}}{{b}} \rangle }
+{\textstyle\frac{2}{3}}\Theta\sigma_{{{a}}{{b}}}
+E_{{{a}}{{b}}}-\D_{\langle {{a}}}\dot{u}_{{{b}}\rangle }
+\sigma_{{c}\langle {{a}}}\sigma_{{{b}}\rangle }{}^{c}+
\omega_{\langle {{a}}}\omega_{{{b}}\rangle} - \dot{u}_{\langle
{{a}}}\dot{u}_{{{b}}\rangle}
\,=\,{\textstyle\frac{1}{2}}\pi^{R}_{{{a}}{{b}}}\;,
\end{equation}
Gravito-electric propagation:
\begin{eqnarray}\label{1+3GrElHO}
 && \dot{E}_{\langle {{a}}{{b}} \rangle }
+\Theta E_{{{a}}{{b}}} -\curl H_{{{a}}{{b}}}
+{\textstyle\frac{1}{2}}(\tilde{\mu}^{m}+\tilde{p}^{m})\sigma_{{{a}}{{b}}}
-2\dot{u}^{c}\ep_{{c}{d}({{a}}}H_{{{b}})}{}^{d} -3\sigma_{{c}\langle
{{a}}}E_{{{b}}\rangle }{}^{c} +\omega^{c}
\ep_{{c}{d}({{a}}}E_{{{b}})}{}^{d}
\nonumber\\&&~~{}=-{\textstyle\frac{1}{2}}(\mu^{R}+p^{R})\sigma_{{{a}}{{b}}}
-{\textstyle\frac{1}{2}}\dot{\pi}^{R}_{\langle {{a}}{{b}}\rangle  }
-{\textstyle\frac{1}{2}}\D_{\langle {{a}}}q^{R}_{{{b}}\rangle }
-{\textstyle\frac{1}{6}}
\Theta\pi^{R}_{{{a}}{{b}}}-{\textstyle\frac{1}{2}}\sigma^{c}{}_{\langle
{{a}}}\pi^{R}_{{{b}}\rangle {c}} -{\textstyle\frac{1}{2}}
\omega^{c}\ep_{{c}({{a}}}^{d}\pi^{R}_{b )d}\;,
\end{eqnarray}
Gravito-magnetic propagation:
\begin{eqnarray}\label{1+3GrMagHO}
 &&\dot{H}_{\langle
{{a}}{{b}} \rangle } +\Theta H_{{{a}}{{b}}} +\curl E_{{{a}}{{b}}}-
3\sigma_{{c}\langle {{a}}}H_{{{b}}\rangle }{}^{c} +\omega^{c}
\ep_{{c}{d}({{a}}}H_{{{b}})}{}^{d}
+2\dot{u}^{c}\ep_{{c}{d}({{a}}}E_{{{b}})}{}^{d}
\nonumber\\&&~~{}={\textstyle\frac{1}{2}}\curl\pi^{R}_{{{a}}{{b}}}-{\textstyle\frac{3}{2}}\omega_{\langle
{{a}}}q^{R}_{{{b}}\rangle
}+{\textstyle\frac{1}{2}}\sigma^{c}{}_{({{a}}}
\ep_{{{b}}){c}}^{\;\;\;\;d}q^{R}_{d}\;,
\end{eqnarray}
Vorticity constraint:
\begin{equation}\label{1+3VorConstrHO}
\D^{{a}}\omega_{{a}} -\dot{u}^{{a}}\omega_{{a}} =0\;,
\end{equation}
Shear constraint:
\begin{equation}\label{1+3ShearConstrHO}
\D^{{b}}\sigma_{{{a}}{{b}}}-\curl\omega_{{a}}
-{\textstyle\frac{2}{3}}\D_{{a}}\Theta +2[\omega,\dot{u}]_{{a}} =
-q^{R}_{a}\;,
\end{equation}
Gravito-magnetic constraint:
\begin{equation}\label{1+3GrMagConstrHO}
 \curl\sigma_{{{a}}{{b}}}+\D_{\langle {{a}}}\omega_{{{b}}\rangle  }
 -H_{{{a}}{{b}}}+2\dot{u}_{\langle {{a}}}
\omega_{{{b}}\rangle  }=0 \;,
\end{equation}
Gravito-electric divergence:
\begin{eqnarray}\label{1+3GrElConstrHO}
&& \D^{{b}} E_{{{a}}{{b}}}
-{\textstyle\frac{1}{3}}\D_{{a}}\tilde{\mu}^{m} -[\sigma,H]_{{a}}
+3H_{{{a}}{{b}}}\omega^{{b}}={\textstyle\frac{1}{2}}\sigma_{{{a}}}^{{b}}q^{R}_{{b}}-
{\textstyle\frac{3}{2}}
[\omega,q^{R}]_{{a}}-{\textstyle\frac{1}{2}}\D^{{b}}\pi^{R}_{{{a}}{{b}}}
 +{\textstyle\frac{1}{3}}\D_{{a}}\mu^{R}
-{\textstyle\frac{1}{3}}\Theta q^{R}_{{a}}\;,
\end{eqnarray}
Gravito-magnetic divergence:
\begin{eqnarray}\label{1+3GrMagDivHO}
 &&\D^{{b}} H_{{{a}}{{b}}}
-(\tilde{\mu}^{m}+\tilde{p}^{m})\omega_{{a}} +[\sigma,E]_{{a}}
 -3E_{{{a}}{{b}}}\omega^{{b}}=-{\textstyle\frac{1}{2}}\curl q^{R}_{{a}}
+(\mu^{R}+p^{R})\omega_{{a}} -{\textstyle\frac{1}{2}}
[\sigma,\pi^{R}]_{{a}} -{\textstyle\frac{1}{2}}\pi^{R}_{{{a}}{{b}}}
\omega^{{b}}\;.
\end{eqnarray}

The standard GR equations are obtained by setting $f(R)=R$ which corresponds to
setting all the right hand sides of these equations to zero. Together with Eqs.
(\ref{eq:cons1})--(\ref{eq:cons4}), these equations govern the
dynamics of the matter and gravitational fields in fourth order
gravity. As we will see the  new source terms in the propagation and constraint
equations will  modify the evolution of the perturbations in a  non-trivial way.

\subsection{Linearized equations }

In the previous section we derived the exact nonlinear equations that govern the exact gravitational
dynamics of fourth order gravity relative to observers comoving with standard matter. These equations
are fully covariant and hold for any spacetime. Consequently, we can linearize these equation
around any chosen background, avoiding the need for choosing coordinates and dealing
directly with physically well defined quantities, rather than metric components
 \footnote{Nevertheless,  our formalism can be easily connected with the
Bardeen or other metric approaches. See  Appendix B for details.}.
These features, which are desirable in the GR case, become essential
for the correct understanding of the evolution of perturbations in
fourth order gravity as well as in other kinds of alternative
gravity theories \cite{ScSante}.

In what follows we will choose a
Friedamnn-Lema\^{\i}tre-Robertson-Walker (FLRW) metric as our
background. We make this choice for a number of different reasons.
First of all the possibility of writing a general fourth order
lagrangian as a simple function of the Ricci scalar is surely
possible for this metric. Secondly, because most of the work in GR
perturbation theory has been performed for this background it makes
a comparison of behavior of GR and fourth order gravity more
straightforward.

The Friedmann background is characterized by the vanishing of all
inhomogeneous and anisotropic quantities $q_a^R, \pi_{ab}^R$ and
defines the order of the quantities appearing in the 1+3 equations
and the linearization procedure. In particular, the quantities that
are zero in the background are considered first-order of in the
linearization scheme. In addition, the Stuart \& Walker lemma ensures
that since these quantities vanish in the background, they are automatically
gauge-invariant \cite{bi:stewart}.

The cosmological equations for the background read:
\begin{eqnarray}\label{f}
&&\Theta^2\,=\,3\tilde{\mu}^{m} + 3\mu^{R}-\frac{\tilde{R}}{2}\;,\\
&&\dot{\Theta}+{\textstyle\frac{1}{3}}\Theta^2
+{\textstyle\frac{1}{2}}(\tilde{\mu}^{m} + 3\tilde{p}^{m})
        +{\textstyle\frac{1}{2}}({\mu}^{R} + 3{p}^{R})=0\;,\\
&& \dot{\mu}^m\,+ \,\Theta\,(\mu^m+{p^m})=0\;,
\end{eqnarray}
where ${\mu}^{R}$ and ${p}^{R}$ are the zero order energy density
and pressure of the curvature fluid, $\tilde{R}$ is the 3-Ricci
scalar and $\tilde{R}=6K/S^2$ with the spatial curvature index
$K=0,\pm1$.

Linearization of the exact propagation and constraint equations about this background then leads to the system:
\begin{eqnarray}
&&\dot{\Theta}+{\textstyle\frac{1}{3}}\Theta^2 -{\tilde\nabla}^{{a}}
A_{{a}}+{\textstyle\frac{1}{2}}(\tilde{\mu}^{m} + 3\tilde{p}^{m})
        \,=\,-{\textstyle\frac{1}{2}}({\mu}^{R} + 3{p}^{R})\label{RayLin}\;,\\
&&\dot{\omega}_{ {{a}} } +2H\omega_{{a}} +{\textstyle\frac{1}{2}}\curl A_{{a}}\,=\,0 \,,\label{pe4l}\\
&& \dot{\sigma}_{ {{a}}{{b}} } +2H\sigma_{{{a}}{{b}}}
        +E_{{{a}}{{b}}}-\D_{\langle {{a}}}A_{{{b}}\rangle }\,=\,-q^{R}_{a}\;,\\
&& \dot{E}_{ {{a}}{{b}}  } +3H E_{{{a}}{{b}}} -\curl H_{{{a}}{{b}}}
        +{\textstyle\frac{1}{2}}(\tilde{\mu}^{m}+\tilde{p}^{m})\sigma_{{{a}}{{b}}}\nonumber\\
        &&~~{}=-{\textstyle\frac{1}{2}}(\mu^{R}+p^{R})\sigma_{{{a}}{{b}}}
        -{\textstyle\frac{1}{2}}\dot{\pi}^{R}_{\langle {{a}}{{b}}\rangle  } -{\textstyle\frac{1}{2}}\D_{\langle
        {{a}}}q^{R}_{{{b}}\rangle } -{\textstyle\frac{1}{6}} \Theta\pi^{R}_{{{a}}{{b}}}\;,\\
&&\dot{H}_{ {{a}}{{b}} } +3H H_{{{a}}{{b}}} +\curl E_{{{a}}{{b}}}\,=\,{\textstyle\frac{1}{2}}\curl\pi^{R}_{{{a}}{{b}}}\;,\\
&&\D^{{b}}\sigma_{{{a}}{{b}}}-\curl\omega_{{a}}-{\textstyle\frac{2}{3}}\D_{{a}}\Theta  =-q^{R}_{a\;,}\\
&&\curl\sigma_{{{a}}{{b}}}+\D_{\langle
        {{a}}}\omega_{{{b}}\rangle}-H_{{{a}}{{b}}}=0 \;,\label{pcc3l}\\
&&\D^{{b}}E_{{{a}}{{b}}}-{\textstyle\frac{1}{3}}\D_{{a}}\tilde{\mu}^{m}
        \,=\,-{\textstyle\frac{1}{2}}\D^{{b}}\pi^{R}_{{{a}}{{b}}}+{\textstyle\frac{1}{3}}\D_{{a}}\mu^{R}
        -{\textstyle\frac{1}{3}}\Theta q^{R}_{{a}}\;,\\
&&\D^{{b}}H_{{{a}}{{b}}}-(\tilde{\mu}^{m}+\tilde{p}^{m})\omega_{{a}}\,=\,-{\textstyle\frac{1}{2}}\curl
        q^{R}_{{a}}
        +(\mu^{R}+p^{R})\omega_{{a}}\;,\label{constrMagLin}\\
&&\D^{{a}}\omega_{{a}}=0\label{VorConstrLin}\;,
\end{eqnarray}
together with the linearized conservation equations
\begin{eqnarray}\label{eq:consLin}
&&\dot{\mu}^m\,=\, - \,\Theta\,(\mu^m+{p^m})\;,\\
&&\3nab^{a}{p^m} =  - (\mu^m+{p^m})\,\udot^{a}\;,\\
&& \dot{\mu^R} + \3nab^{a}q^R_{a} = -
\,\Th\,(\mu^R+p^R) +\mu^{m}\frac{f''\,\dot{R}}{f'^{2}}\;,\\
&& \dot{q}^R_{\lgl a\rgl} + \3nab_{a}p^R + \3nab^{b}\pi^R_{ab} = -
\,{\textstyle\frac{4}{3}}\,\Th\,q^R_{a}  - (\mu^R+p^R)\,\udot_{a}
+\mu^{m}\frac{f''\,\D_{a}{R}}{f'^{2}} \;,
\end{eqnarray}
obtained from (\ref{eq:cons1})--(\ref{eq:cons4}). Note that at first
order the equation of the vorticity \rf{VorConstrLin} is homogeneous
i.e. the evolution of the vorticity is decoupled. This will be
important in the next section when we will derive the perturbations
equations. These equations provide the basis for a covariant and
gauge-invariant description of perturbations of $f(R)$ theories of
gravity.
\section{Dynamics of scalar Perturbations}
\subsection{Perturbation Equations}
We are now ready to analyze the evolution of the density
perturbations on a FLRW background. The quantities appearing in the
linearized equations given in the previous section can be decomposed
in scalar vector and tensor components, i.e.
\begin{equation}
    V_a= \bar{V}_a + \hat{V}_a=\eta^{abc}\3nab_{b}\bar{V}_{c}+\3nab^a
    \hat{v}\,,\qquad
     \mbox{where} \qquad  \3nab^a \bar{V}_a =0\;, \qquad
     \eta^{abc}\3nab_{b}\hat{V}_{c}=0\;,
\end{equation}
and
\begin{equation}
    W_{ab}=
     \bar{W}_{ab}+\hat{W}_{ab}+W^{*}_{ab}= \bar{W}_{ab} +
    \3nab_a\bar{W}_b +  \3nab_a\3nab_b W^{*}\;,
\end{equation}
where
\begin{equation}
    \3nab^a \bar{W}_{ab} =0\;, \qquad (\c\,\hat{W})_{ab} =0\;,
    \qquad (\c\,W^{*})_{ab}=0\;,
\end{equation}
and both of these decompositions are unique. Note that here we
define scalars, vectors or tensors as quantities that transform like
scalars, solenoidal vectors or symmetric tensors, or are obtained
from them using the $h_{ab}$ or $\3nab_{a}$ operators
\cite{EBH}.

In linear regime and in homogeneous and isotropic backgrounds these
different components do not interact with each other. In the
following we will focus only on the evolution of scalar
perturbations because they are directly related with density
fluctuations. This can be done simply discarding the non scalar
quantities in the equation above i.e. setting
\begin{equation}
V_{{a}}=\D_{{a}} V\,,~~W_{{{a}}{{b}}}= \D_{\langle{{a}}}\D_{{{b}}\rangle}W\,.
\end{equation}
The identities in Appendix \ref{CovID}, the vorticity constraint
equation (\ref{1+3VorConstrHO}) and the gravito-magnetic constraint
equation (\ref{1+3GrMagConstrHO}) then show that
\begin{equation}\label{s1}
\curl V_{{a}}=0=\curl W_{{{a}}{{b}}}\,,~ \D^{{b}} W_{{{a}}{{b}}}={\textstyle\frac{2}{3}}\D^2( \D_{{a}}
W)\,,~\omega_{{a}}=0=H_{{{a}}{{b}}}\,,
\end{equation}
as in standard General Relativity.

In order to derive the equations governing density perturbations in
the general case, we define the density and expansion gradients
\begin{equation}\label{s3}
{\cal D}^{m}_{{a}}=\frac{S}{\mu^{m}}\D_{{a}}\mu^{m}\,,\qquad
Z_{{a}}=S\D_{{a}}\Theta\,,\qquad C_{{a}}=S\D_{{a}}\tilde{R}\;,
\end{equation}
and the (dimensionless) gradients describing inhomogeneity in the Ricci scalar:
\begin{equation}\label{s4}
{\cal R}_{{a}}=S\D_{{a}} R\,,\qquad\Re_a=S\D_{{a}} \dot{R}\;.
\end{equation}
 Another important quantity in the treatment of the evolution of
the density perturbations is the Newtonian potential (defined through the divergence of the
electric part of the Weyl tensor (\ref{1+3GrElConstrHO}) \cite{EBH}).
\begin{eqnarray}\label{eq:BardeenPHI}
   \nonumber &&\Phi_{a}^{N}=S^{2} \mu^{tot}\;  \dd_{a}^{tot}
    \\&& \nonumber = \frac{2  S^{2}  \mu  \Theta }{2\Theta f'+3 \dot{R}f''}\;
    \dd_{a}^{m}
    +\frac{3  \dot{R} f''}{2 \left(2 \Theta  f'+3 \dot{R}
    f''\right)}\;C_a
    -\frac{2   S^{2} f'' \Theta ^2}{2 \Theta f'+3\dot{R}
    f''}\;\Re_a\\&&
    +\frac{ S^{2} \left[f'' \left(f-2 \mu +2 \dot{R} \Theta f''\right)
    -2 \dot{R} \Theta  f' f^{(3)}\right]\Theta }{ f' \left(2 \Theta  f'+3
   \dot{R} f''\right)}\;\mathcal{R}_a
\end{eqnarray}
where $\dd_{a}^{tot}$ represents the total energy density
fluctuation.

 Using equations (\ref{RayLin}) -(\ref{constrMagLin}), equations
(\ref{eq:cons1})--(\ref{eq:cons4}), the identities in Appendix
\ref{CovID}, assuming matter to be a barotropic perfect fluid with
barotropic factor $w=p^m/\mu^{m}$ and that the vorticity is zero
\footnote{This last assumption does not compromise the generality of
our treatment because it is well known that although the 3-gradient
of the vorticity acts as a source term for the evolution equation of
$\dd_a$ it does not affect the scalar part of this quantity  \cite{EBH}.}, we obtain the following
system of evolution equations for the above variables:
\begin{eqnarray}
\dot{{\cal D}}^m_{{a}} &=&w\Theta{\cal D}^m_{{a}}-(1+w)Z_{{a}}\,,\label{s5}\\
\nn\dot{Z}_{{a}} &=& \left(\frac{\dot{R}
   f''}{f'}-\frac{2 \Theta }{3}\right)Z_a+
   \left[\frac{3 (w -1) (3 w +2)}{6 (w +1)} \frac{\mu}{ f'} + \frac{2 w \Theta ^2
   +3 w (\mu^{R}+3  p^{R}) }{6 (w +1) }\right]
   {\cal D}^m_{a}+\frac{\Theta f''}{f'}\Re_{a}\\&&+
   \left[\frac{1}{2}-\frac{ f''}{f'} \frac{K}{S^2}-\frac{1}{2} \frac{f}{f'}\frac{ f''}{f'}- \frac{f''}{f'} \frac{\mu}{
   f'} + \dot{R} \Theta  \left(\frac{f''}{f'}\right)^{2}+ \dot{R} \Theta \frac{ f^{(3)}}{ f'}\right]\mathcal{R}_a
   -\frac{w}{w +1} \3nab^{2}{\cal
   D}^m_{a}-\frac{ f''}{f'}\3nab^{2}\mathcal{R}_{a}\,,\\
\dot{{\cal R}}_a&=&\Re_{a}-\frac{w }{w +1}\dot{R}\;{\cal D}^m_{{a}}\,,\label{eqZa}\\
\nn\dot{\Re}_a&=&- \left(\Theta + 2\dot{R} \frac{
   f^{(3)}}{f''}\right)\Re_{a}- \dot{R} Z_{a} -
   \left[\frac{ (3 w -1)}{3} \frac{\mu}{f''} + 3\frac{w}{w +1}
   (p^{R}+\mu^{R}) \frac{f'}{f''}+ \frac{w}{3(w +1)} \dot{R} \left(\Theta
   -3 \dot{R} \frac{f^{(3)}}{f''}\right)\right]{\cal D}^m_{{a}}\\&&\nn\left[3\frac{K}{S^2}-\left(
   \frac{1}{3}\frac{f'}{f''}+\frac{f^{(4)}}{f'} \dot{R}^2+\Theta  \frac{f^{(3)}}{f'}
   \dot{R}-\frac{2}{9} \Theta ^2 +\frac{1}{3}(\mu^{R}+3 p^{R}) + \ddot{R} \frac{f^{(3)}}{f''}
   - \frac{1}{6}\frac{f}{f'}+\frac{1}{2} (w +1) \frac{\mu}{f'} -\frac{1}{3} \dot{R} \Theta
 \frac{f''}{f'}\right)\right]\mathcal{R}_{a}\\&&+\3nab^{2}\mathcal{R}_{a}\,,
\end{eqnarray}
together with the constraint
\begin{equation}\label{Gauss}
  \frac{C_a}{S^2}+ \left(\frac{4  }{3}\Theta +\frac{2 \dot{R}
   f''}{f'}\right) Z_a-2\frac{
    \mu }{f'}{\cal D}^m_{{a}}+ \left[2 \dot{R}
   \Theta  \frac{f^{(3)}}{ f'}-\frac{f''}{ f'} \left(f-2 \mu +2
   \dot{R} \Theta  f''+2\frac{K}{S^{2}}\right)\right]\mathcal{R}_a+\frac{2 \Theta
   f''}{f'}\Re_a-\frac{2 f''}{f'}\3nab^{2}\mathcal{R}_a=0\,.
\end{equation}
The propagation equation for the variable C is
\begin{eqnarray}
&& \dot{C}_a=\nonumber K^2 \left[\frac{36  f'' \mathcal{R}_a}{S^2
    \left(2 \Theta
    f'+3 \dot{R} f''\right)}-\frac{36
   f' \dd^{m}_a}{S^2 \left(2 \Theta  f'+3 \dot{R}
   f''\right)} \right] +K\left\{\frac{6  f'}{S^2 \left(2
   \Theta  f'+3 \dot{R} f''\right)}C_a \right.\\&&\left.+\dd^m_a
   \left(\frac{16 \omega  \Theta }{3(\omega +1)}-\frac{ 4f'\Theta ^2-12f'
   \mu^{R}}{2 \Theta  f'+3 \dot{R}
   f''}\right)\nonumber-\frac{12  f''}{2 \Theta
    f'+3 \dot{R} f''}\3nab^{2}\mathcal{R}_a+\left(\frac{12 \Theta
   f''}{2 \Theta  f'+3 \dot{R}
   f''}+2\frac{f''}{f'}\right)\Re_a\right.\\&&\left.\nonumber+\left[-\frac{2 S^2 \left(\Theta
    f''-3 \dot{R} f^{(3)}\right)}{3 f'}\frac{ 12 \dot{R} \Theta  f'
   f^{(3)}-2 f'' \left(3 f-2 \left(\Theta ^2-3 \mu^{R}\right) f'+6 \dot{R} \Theta
   f''\right)}{\left(2 \Theta  f'+3 \dot{R}
   f''\right) f'}\right]\mathcal{R}_a\right\} \\&&+\3nab^{2}\left[\frac{4 \omega
    S^2 \Theta }{3 (\omega +1)}\dd^m_a+\frac{2  S^2
   f''}{f'}\Re_a-\frac{2 S^2 \left(\Theta
    f''-3 \dot{R} f^{(3)}\right)}{3 f'}\mathcal{R}_a\right]\,,
\end{eqnarray}
this equation, which is redundant, will be used in Section
\ref{esempioRn} to substitute \rf{eqZa} because of its specific form
in the long wavelength limit \cite{Conserved}.

\subsection{Scalar Variables}
The variables we have defined above describe the general evolution
of the density perturbations and the other scalars on a FLRW
background. The phenomenon of the clustering of matter is
traditionally  described, however, considering only the scalar part
of these variables. This can be easily done using the {\em local}
decomposition \cite{EBH}
\begin{equation}\label{localdecomposition}
   S\3nab^{a}X_a=X_{ab}= \frac{1}{3}h_{ab}X+\Sigma^{X}_{ab}+X_{[ab]}
   \qquad\mbox{where}\qquad
   \Sigma^{X}_{ab}=X_{(ab)}-\frac{1}{3}h_{ab}X\;.
\end{equation}
so that the operator $\3nab_{a}$ applied to the \rf{s3} and \rf{s4}
extracts the scalar part of the perturbation variables. In this way
we can define the scalar quantities
\begin{equation}\label{ScaVar}
\Delta_{m}=S\3nab^{a}{\cal D}^{m}_{{a}}\,,\quad
Z=S\3nab^{a}Z_{{a}}\,,\quad C=S\3nab^{a}C_{{a}}\,,\quad{\mathcal
R}=S\3nab^{a}{\mathcal R}_{{a}}\,,\quad\Re=S\3nab^{a}\Re_a\quad
\Phi^{N}=S\3nab^{a}\Phi_{a}^{N}\,.
\end{equation}
which will characterize the evolution of the spherically symmetric
part of the gradients (\ref{s3}-\ref{s4}). The evolution equations
for the first four of these variables are
\begin{eqnarray}
&&\dot{\Delta}_m =w\Theta \Delta_m-(1+w)Z\,,\label{eqDelta}\\
&&\nn\dot{Z} = \left(\frac{\dot{R}
   f''}{f'}-\frac{2 \Theta }{3}\right)Z+
   \left[\frac{3 (w -1) (3 w +2)}{6 (w +1)} \frac{\mu}{ f'} + \frac{2 w \Theta ^2
   +3 w (\mu^{R}+3  p^{R}) }{6 (w +1) }\right]
   \Delta_m+\frac{\Theta f''}{f'}\Re\\&&+
   \left[\frac{1}{2}-\frac{1}{2} \frac{f}{f'}\frac{ f''}{f'}- \frac{f''}{f'} \frac{\mu}{
   f'} + \dot{R} \Theta  \left(\frac{f''}{f'}\right)^{2}+ \dot{R} \Theta \frac{ f^{(3)}}{ f'}\right]\mathcal{R}
   -\frac{w}{w +1} \3nab^{2}{\Delta}_m-\frac{ f''}{f'}\3nab^{2}\mathcal{R}\,,\\
&&\dot{{\cal R}}=\Re-\frac{w }{w +1}\dot{R}\;{\Delta}_m\,,\\
&&\nn\dot{\Re}=- \left(\Theta + 2\dot{R} \frac{
   f^{(3)}}{f''}\right)\Re- \dot{R} Z -
   \left[\frac{ (3 w -1)}{3} \frac{\mu}{f''} + 3\frac{w}{w +1}
   (p^{R}+\mu^{R}) \frac{f'}{f''}+ \frac{w}{3(w +1)} \dot{R} \left(\Theta
   -3 \dot{R} \frac{f^{(3)}}{f''}\right)\right]{\Delta}_m\\&&\nn+\left[2 \frac{K}{S^2}-\left(
   \frac{1}{3}\frac{f'}{f''}+\frac{f^{(4)}}{f'} \dot{R}^2+\Theta  \frac{f^{(3)}}{f'}
   \dot{R}-\frac{2}{9} \Theta ^2 +\frac{1}{3}(\mu^{R}+3 p^{R}) + \ddot{R} \frac{f^{(3)}}{f''}
   - \frac{1}{6}\frac{f}{f'}+\frac{1}{2} (w +1) \frac{\mu}{f'} -\frac{1}{3} \dot{R} \Theta
   \frac{f''}{f'}\right)\right]\mathcal{R}\\&&+\3nab^{2}\mathcal{R}\,,\label{eqRho}
\end{eqnarray}
\begin{eqnarray}
&& \dot{C}=\nonumber K^2 \left[\frac{36  f'' \mathcal{R}}{S^2
    \left(2 \Theta
    f'+3 \dot{R} f''\right)}-\frac{36
   f' \Delta}{S^2 \left(2 \Theta  f'+3 \dot{R}
   f''\right)} \right] +K\left[\frac{6  f'}{S^2 \left(2
   \Theta  f'+3 \dot{R} f''\right)}C +\Delta
   \left(\frac{4 \omega  \Theta }{\omega +1}-\frac{ 4f'\Theta ^2-12f'
   \mu^{R}}{2 \Theta  f'+3 \dot{R}
   f''}\right)\right.\\&&\nonumber\left.-\frac{12  f''}{2 \Theta
    f'+3 \dot{R} f''}\3nab^{2}\mathcal{R}+\frac{12 \Theta
   f''}{2 \Theta  f'+3 \dot{R}
   f''}\Re+\frac{ 12 \dot{R} \Theta  f'
   f^{(3)}-2 f'' \left(3 f-2 \left(\Theta ^2-3 \mu^{R}\right) f'+6 \dot{R} \Theta
   f''\right)}{\left(2 \Theta  f'+3 \dot{R}
   f''\right) f'}\mathcal{R}\right] \\&&+\3nab^{2}\left[\frac{4 \omega
    S^2 \Theta }{3 (\omega +1)}\Delta+\frac{2  S^2
   f''}{f'}\Re-\frac{2 S^2 \left(\Theta
    f''-3 \dot{R} f^{(3)}\right)}{3
    f'}\mathcal{R}\right]\,,\label{eqC}
\end{eqnarray}
together with the constraint
\begin{equation}\label{Gauss1}
  \frac{C}{S^2}+ \left(\frac{4  }{3}\Theta +\frac{2 \dot{R}
   f''}{f'}\right) Z-2\frac{
    \mu }{f'}{\Delta}_m+ \left[2 \dot{R}
   \Theta  \frac{f^{(3)}}{ f'}-\frac{f''}{ f'} \left(f-2 \mu +2
   \dot{R} \Theta  f''\right)\right]\mathcal{R}+\frac{2 \Theta
   f''}{f'}\Re-\frac{2 f''}{f'}\3nab^{2}\mathcal{R}=0\,.
\end{equation}
In standard GR, only the first two equations  and the last one are
present and the density perturbations are governed by a second-order
equation for $\Delta^{m}$ whose independent solutions are adiabatic
growing and decaying modes. The presence of fourth order corrections
introduces important changes to this  picture. In fact, in this case
the evolution of the density perturbations is described by a closed
{\em fourth} order differential equation which can be obtained form
the above first order equations. This follows clearly from our two
effective fluids interpretation.
\subsection{Harmonic analysis}
The system (\ref{eqDelta})-(\ref{eqRho}) is a system of four partial
differential equations which is far too complicated to be solved
directly. For this reason, following a standard procedure we perform
an harmonic decomposition. This allows one to reduce equations
(\ref{eqDelta})-(\ref{eqRho}) to ordinary differential equations
which are somewhat easier to solve.

In the covariant approach the harmonic decomposition is performed
using the trace-free symmetric tensor eigenfunctions of the spatial
the Laplace-Beltrami operator defined by \cite{BDE}:
\begin{eqnarray}\label{eq:harmonic}
  \3nab^{2}Q = -\frac{k^{2}}{a^{2}}Q\;,
\end{eqnarray}
where $k=2\pi S/\lambda$ is the wavenumber and $\dot{Q}=0$. Using
these harmonics we can expand every first order quantity in the
equations above \footnote{Note that the underlying assumption in
this decomposition is that the perturbation variables can be
factorized into purely temporal and purely spatial components.},
\begin{equation}\label{eq:developmentdelta}
X(t,\mathbf{x})=\sum X^{(k)}(t)\;Q^{(k)}(\mathbf{x})
\end{equation}
where $\sum$ stands for both a summation over a discrete index or an
integration over a continuous one.

Developing \rf{ScaVar} in terms of $Q$,
(\ref{eqDelta})-(\ref{Gauss}) reduce to
\begin{eqnarray}
\dot{\Delta}_{m}^{(k)} &=&w\Theta \Delta_{m}^{(k)}-(1+w)Z^{(k)}\,,\label{eqDeltaHarm}\\
\dot{Z}^{(k)} &=& \left(\frac{\dot{R}
   f''}{f'}-\frac{2 \Theta }{3}\right)Z^{(k)}+
   \left[\frac{3 (w -1) (3 w +2)}{6 (w +1)} \frac{\mu}{ f'} + \frac{2 w \Theta ^2
   +3 w (\mu^{R}+3  p^{R}) }{6 (w +1) }-\frac{w}{w +1} \frac{k}{S^2}\right]
   \Delta_{m}^{(k)}+\frac{\Theta f''}{f'}\Re^{(k)}\nonumber \\&&+
   \left[\frac{1}{2}-\frac{ f''}{f'} \frac{k}{S^2}-\frac{1}{2} \frac{f}{f'}\frac{ f''}{f'}- \frac{f''}{f'} \frac{\mu}{
   f'} + \dot{R} \Theta  \left(\frac{f''}{f'}\right)^{2}+ \dot{R} \Theta \frac{ f^{(3)}}{ f'}\right]\mathcal{R}^{(k)}\,,\\
\dot{{\cal R}}^{(k)}&=&\Re^{(k)}-\frac{w }{w +1}\dot{R}\;{\Delta}_{m}^{(k)}\,,\label{eqZHarm}\\
\dot{\Re}^{(k)}&=&- \left(\Theta + 2\dot{R} \frac{
   f^{(3)}}{f''}\right)\Re^{(k)}- \dot{R} Z^{(k)} -
   \left[\frac{ (3 w -1)}{3} \frac{\mu}{f''} + 3\frac{w}{w +1}
   (p^{R}+\mu^{R}) \frac{f'}{f''}+ \frac{w}{3(w +1)} \dot{R} \left(\Theta
   -3 \dot{R} \frac{f^{(3)}}{f''}\right)\right]{\Delta}_{m}^{(k)}\nonumber\\ &&+\left[\frac{k}{S^2}+2 \frac{K}{S^2}-\left(
   \frac{1}{3}\frac{f'}{f''}+\frac{f^{(4)}}{f'} \dot{R}^2+\Theta  \frac{f^{(3)}}{f'}
   \dot{R}-\frac{2}{9} \Theta ^2\right.\right.
   \nn\\&&  \left.\left.+\frac{1}{3}(\mu^{R}+3 p^{R}) + \ddot{R} \frac{f^{(3)}}{f''}
   - \frac{1}{6}\frac{f}{f'}+\frac{1}{2} (w +1) \frac{\mu}{f'} -\frac{1}{3} \dot{R} \Theta
 \frac{f''}{f'}\right)\right]\mathcal{R}^{(k)}\,,\label{eqRho2}
\\
\dot{C}^{(k)}&=& K^2 \left[\frac{36  f''
\mathcal{R}^{(k)}}{S^2
    \left(2 \Theta
    f'+3 \dot{R} f''\right)}-\frac{36
   f' \Delta^{(k)}}{S^2 \left(2 \Theta  f'+3 \dot{R}
   f''\right)}-\frac{12  f''}{S^{2}\left(2 \Theta
    f'+3 \dot{R} f''\right)}\mathcal{R}^{(k)} \right]\nn\\&& +K\left[\frac{6  f'}{S^2 \left(2
   \Theta  f'+3 \dot{R} f''\right)}C^{(k)} +
   \left(\frac{4 \omega  \Theta }{\omega +1}-\frac{ 4f'\Theta ^2-12f'
   \mu^{R}}{2 \Theta  f'+3 \dot{R}
   f''}\right)\;\Delta^{(k)}\right.\nonumber \\&&\left.+\frac{12 \Theta
   f''}{2 \Theta  f'+3 \dot{R}
   f''}\Re^{(k)}+\frac{ 12 \dot{R} \Theta  f'
   f^{(3)}-2 f'' \left(3 f-2 \left(\Theta ^2-3 \mu^{R}\right) f'+6 \dot{R} \Theta
   f''\right)}{\left(2 \Theta  f'+3 \dot{R}
   f''\right) f'}\mathcal{R}^{(k)}\right] \nn\\&&+\frac{k}{S^{2}}\left[\frac{4 \omega
    S^2 \Theta }{3 (\omega +1)}\Delta^{(k)}+\frac{2  S^2
   f''}{f'}\Re^{(k)}-\frac{2 S^2 \left(\Theta
    f''-3 \dot{R} f^{(3)}\right)}{3
    f'}\mathcal{R}^{(k)}\right]\,,\label{eqCHarm}
    \end{eqnarray}
\begin{eqnarray}
 0&=& \frac{C^{(k)}}{S^2}+ \left(\frac{4  }{3}\Theta +\frac{2
\dot{R}
   f''}{f'}\right) Z^{(k)}-2\frac{
    \mu }{f'}{\Delta}_m^{(k)}+ \left[2 \dot{R}
   \Theta  \frac{f^{(3)}}{ f'}-\frac{f''}{ f'} \left(f-2 \mu +2
   \dot{R} \Theta  f''\right)-2 \frac{ f''}{f'} \frac{k}{S^2}\right]\mathcal{R}^{(k)}+\frac{2 \Theta
   f''}{f'}\Re^{(k)}\;.\nn\\\label{constrCZDel}
\end{eqnarray}
Finally it is useful to write equations (\ref{eqDeltaHarm}-\ref{eqRho2}) as a pair of second order equation.
In this way the GR limit is more transparent when the written in this form. They are:
\begin{eqnarray}
   &&\nn\ddot{\Delta}^{(k)}-\left[\left(\omega
   -\frac{2}{3}\right) \Theta +\frac{\dot{R}
   f''}{f'}\right] \dot{\Delta}^{(k)}-\left[\omega  k^2-\omega  (3
   p^{R}+\mu^{R})-\frac{2 \omega  \dot{R} \Theta
   f''}{f'}-\frac{\left(3 \omega ^2-1\right) \mu }{f'}\right]\Delta^{(k)}\\&&=
   \frac{1}{2}(w +1)\left[2 \frac{k^2}{S^2}f''+
   \left(f-2 \mu +2 \dot{R} \Theta  f''\right)\frac{f''}{f'^2}
   -2  \dot{R} \Theta
   \frac{f^{(3)}}{f'}\right] \mathcal{R}^{(k)} -\frac{(w +1) \Theta
   f'' }{f'}\dot{\mathcal{R}}^{(k)}\\&&
   \nn f''\ddot{\mathcal{R}}^{(k)}+\left(\Theta f'' +2 \dot{R}
   f^{(3)}\right)
   \dot{\mathcal{R}}^{(k)}-\left[\frac{k^2}{S^2}f''+ 2 \frac{K}{S^2}f''
  +\frac{2}{9} \Theta^2 f''- (w +1) \frac{\mu}{2 f'}f''- \frac{1}{6}(\mu^{R}+ 3
p^{R})f''\right.\\&&\nn\left.-\frac{f'}{3}+ \frac{f}{6 f'}f'' +
\dot{R} \Theta  \frac{f''^{2}}{6 f'} -
   \ddot{R} f^{(3)}- \Theta f^{(3)} \dot{R}- f^{(4)}\dot{R}^2
   \right]\mathcal{R}^{(k)}=-
   \left[ \frac{1}{3}(3 w -1) \mu \right.\\&&\left.+\frac{w}{1+w} \left(f^{(3)}
   \dot{R}^2+  (p^{R}+\mu^{R}) f'+ {\textstyle\frac{7}{3}}\dot{R} \Theta f''
 +\ddot{R} f'' \right)\right]\Delta^{(k)}-\frac{(w -1) \dot{R} f''}{w +1} \dot{\Delta}^{(k)}
\end{eqnarray}
In the GR limit we have $f=R$, so the above equations reduce to
\begin{eqnarray}
 &&\nn\ddot{\Delta}^{(k)}-\left(\omega-{\textstyle\frac{2}{3}}\right) \Theta\dot{\Delta}^{(k)}
 -\left[\omega k^2-\left({\textstyle\frac{1}{2}}+\omega-{\textstyle\frac{3}{2}}w^2\right)\mu\right]\Delta^{(k)}=0\;,\\
&&\mathcal{R}=\left(3\omega-1\right)\mu\Delta^{(k)}\;.
\end{eqnarray}
The second of these equations is just the spatial Laplacian of the
trace of the Einstein Field equations $R=3p-\mu$.
\section{Example: $R^n$-gravity}\label{esempioRn}
Let us now apply the equations derived in the above sections to the
simplest example of fourth order theory of gravity: $R^n$-gravity.
In this theory $f(R)=\chi R^{n}$ and the action  reads
\begin{equation}\label{71-curv1}
{\cal A}=\int d^4x \sqrt{-g} \left[\chi R^{n}+{L}_{M} \right]\;,
 \end{equation}
where $\chi$ a the coupling constant with suitable dimensions and
$\chi=1$ for $n=1$.

If $R\neq 0$ the field equations for this theory read
\begin{eqnarray}\label{equazioni di campo Rn}
G_{ab}=\chi^{-1}\frac{\tilde{T}_{ab}^{M}}{nR^{n-1}}+T^{R}_{ab}
\end{eqnarray}
where
\begin{eqnarray}
\tilde{T}_{ab}^{M}&=&\chi^{-1}\frac{T_{ab}^{M}}{nR^{n-1}}\,,\\
T^{R}_{ab}&=&(n-1)\left\{-\frac{R}{2 n}g_{ab} +\left[\frac{R^{;c
d}}{R}+(n-2)\frac{R^{;c} R^{;d}}{R^{2}}\right](g_{c a}g_{d b}-g_{c
d}g_{ab})\right\}\,.
\end{eqnarray}
The FLRW dynamics of this model has been investigated via a complete
phase space analysis in \cite{cdct:dynsys05}. This analisys shows
that for specific intervals of the parameter $n$ there is a set of
initial conditions with non zero measure for which the cosmic
histories include a transient decelerated phase which evolves
towards an accelerated expansion one. This first phase was argued to
be suitable for the structure formation to take place.

In what follows we will analyze the evolution of the scalar
perturbations during this phase in the long wavelength limit. In
this approximation the wavenumber $k$ is considered to be so small
that the wavelength $\lambda=2\pi S/k$ associated with it is much
larger than the Hubble radius. Equation \rf{eq:harmonic} then
implies that all the Laplacians can be neglected and the spatial
dependence of the perturbation variables can be factored out. It is
also well known \cite{Conserved} that in this limit and in spatially
flat ($K=0$) backgrounds the \rf{eqCHarm} reduces to $\dot{C}=0$
i.e. the variable $C$ is conserved so that the number of
perturbations equations can be reduced to three.

\begin{figure}
  \includegraphics[width=11cm]{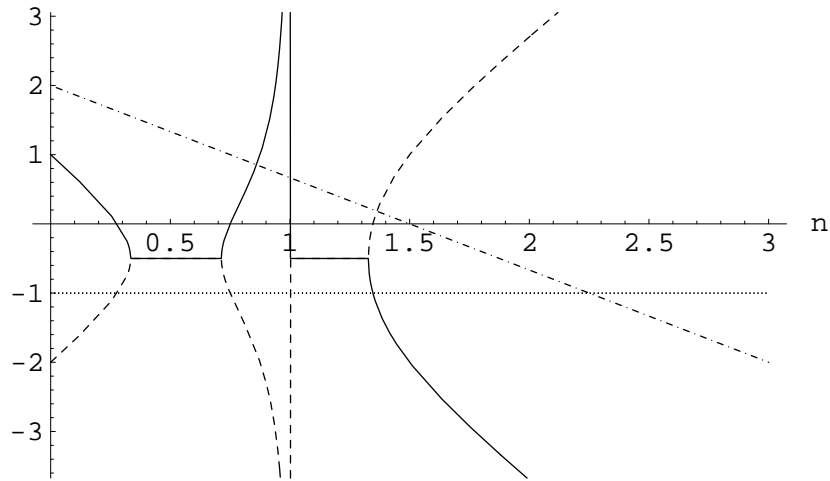}\\
  \caption{Plot of the real part of  the exponents of each modes of the solution
  \rf{soluzione delta gen LWL} against $n$ in the case of dust ($w=0$). The
  continuous and dashed line represent the modes $t^{\alpha_{\pm}}$
  respectively (note how they coincide when $\alpha_{\pm}$ are complex),
  the dashed-dot line represents the mode $ t^{2-\frac{4 n}{3
  (\omega+1)}}$ and the dot line the mode $t^{-1}$.}\label{plotdust}
\end{figure}

Let us now set the background to be the  transient solution
\begin{equation}
    S=S_0 t^{\frac{2n}{3(1+w)}}\;,\qquad k=0\;,\qquad \mu=\mu_{0} t^{-2n}
\end{equation}
of \cite{cdct:dynsys05}. The expansion, the Ricci scalar, the
curvature fluid  pressure, the curvature fluid energy density and
the effective matter energy density take the form:
\begin{eqnarray}
        \Theta &=&\frac{2 n}{t (\omega +1)}\,,\\
          R &=& \frac{4 n [4 n-3 (\omega +1)]}{3 t^2 (\omega +1)^2}\,, \\
          \mu^{R} &=& \frac{2 (n-1) [2 n (3 \omega +5)-3 (\omega +1)]}{3 t^2 (\omega +1)^2}\,, \\
          p^{R} &=& \frac{2 (n-1) \left[n \left(6 \omega ^2+8 \omega -2\right)-3 \omega  (\omega
   +1)\right]}{3 t^2 (\omega +1)^2}\,,\\
   \mu &=& \left(\frac{3}{4}\right)^{1-n} n \chi  \left(\frac{n (4 n-3 (\omega +1))}{t^2
   (\omega +1)^2}\right)^{n-1} \frac{4 n^2-2 (n-1) [2 n (3 \omega +5)-3 (\omega +1)]}{3 (\omega +1)^2
   t^2}\,.
        \end{eqnarray}
Substituting in the equations given above and passing to the  long
wavelength limit we obtain
\begin{eqnarray}\label{sistemLWLBckgr}
\nn \dot{\Delta}_m &=& \left[-\frac{2 n}{w
   +1}-\frac{6 (n-1) n}{n+3 (n-1) w -3}+1\right] \frac{\Delta_m}{t}
   -\frac{3 (w +1)^2 }{4 a_0^2 [n+3 (n-1)w -3] } \;C_0 \;t^{1-\frac{4 n}{3 (w +1)}}\\&&
   +\frac{3 (n-1)(w +1)^2 [n (6 w +8)-15 (w +1)]}{4 [n+3 (n-1) w -3] [4 n-3 (w +1)]}\;\mathcal{R}\;t
   -\frac{9 (n-1) (w+1)^3  t^2}{4 [n+3 (n-1) w -3] [4 n-3 (w +1)]}\;\Re
   \;t^2\;,\\
\dot{\mathcal{R}} &=& \Re+\frac{8 n w  (4 n-3 (w +1))}{3  (w +1)^3} \frac{ \Delta_m }{t^3 }\;,  \\
\nn\dot{\Re} &=& \nonumber\frac{2 n (4 n-3 w -3) }{(w +1) (n+3 (n-1)
w
          -3)}\frac{C_0}{a_0^2} t^{-\frac{4 n}{3 (w +1)}-2}+2 \left(\frac{3n
          (n-1)}{n+3 (n-1) w -3}-\frac{n}{w +1}+2n-4\right)
          \frac{\Re}{t}\\&&+2 \left( -\frac{9 n (n-2) (n-1)}{n+3(n-1) w -3}-2 n^{2}+7 n+\frac{3 n^{2} (9
          n-26)+57}{9 (w +1) (n-1)}+\frac{8 n^{2} (n-2) }{9 (w +1)^2 (n-1)}-6\right)\frac{\mathcal{R}}{t^2}\\&&
          \nonumber+\frac{16 n (4 n+3 (n-1) w -3) (4 n-3 (w +1)) \left((9 w  (w +1)+8) n^2-(3 w
         (9 w +8)+13) n+3 (w +1) (6 w +1)\right)}{27 (n-1)(w +1)^4 (n+3 (n-1) w -3) } t^4\, \Delta_m\;,
\end{eqnarray}
where $C_0$ is the conserved value for the quantity $C$. The evolution of density
perturbations can then be decoupled via the third order equation
\begin{eqnarray}\label{EqDelta3Ord}
(n-1)\dddot{\Delta}_m-(n-1)\left(\frac{4 n \omega }{\omega
+1}-5\right)\frac{
   \ddot{\Delta}_m
   }{t}+ \mathcal{D}_1 (n,w)\;\frac{\dot{\Delta}_m}{t^2}+\mathcal{D}_2 (n,w)\frac{\Delta _m}{ t^3}+\mathcal{D}_3(n,w)\;\mathcal{C}_0\; t^{-\frac{4 n}{3 (\omega +1)}-1}=0
\end{eqnarray}
where
\begin{eqnarray}
  \mathcal{D}_1  (n) &=& -\frac{2 \left(-9 (2 (n-1) n+1) \omega ^2+6 n (n (4 n-7)+1)
   \omega +18 \omega +n (4 n (8 n-19)+33)+9\right) }{9 (\omega
   +1)^2} \\
\mathcal{D}_2(n) &=&\frac{2 ((2 n-1) \omega -1) (4 n-3 (\omega +1))
(3 (\omega +1)+n
   (-9 \omega +n (6 \omega +8)-13)) }{9  (\omega +1)^3
  }\\
\mathcal{D}_3(n) &=& -\frac{n (21 \omega -6 n (\omega +2)+31)-18
(\omega +1)
   }{6  a_0^2}\;
\end{eqnarray}
 This equation admits the general solution
\begin{eqnarray}\label{soluzione delta gen LWL}
   &&\Delta_m=K_1 t^{\frac{2 n \omega
   }{\omega +1}-1}+K_2 t^{\alpha _+}+K_3 t^{\alpha _-}-K_4 \frac{\mathcal{C}_0}{a_0^2}
   t^{2-\frac{4 n}{3 (\omega+1)}}\;,
\end{eqnarray}
where
\begin{eqnarray}
 &&\alpha_{\pm} = -\frac{1}{2}+\frac{n \omega }{\omega +1}\pm \frac{\sqrt{(n-1)  \left(4 (3
   \omega +8)^2 n^3-4 (3 \omega  (18 \omega +55)+152) n^2+3 (\omega +1) (87
   \omega +139) n-81 (\omega +1)\right)}}{6 (n-1) (\omega
   +1)^2}\nonumber\\\\
 &&K_4= \frac{9 (\omega +1)^3 (18 (\omega +1)+n (-21 \omega +6 n (\omega +2)-31))}{8 (n
   (6 \omega +4)-9 (\omega +1)) \left(6 (\omega +2) n^3-(9 \omega +19) n^2-3
   (\omega +1) (3 \omega +1) n+9 (\omega +1)^2\right)}\;.
\end{eqnarray}
Let us now focus on the case of dust ($w=0$). The above solution
becomes
\begin{eqnarray}\label{soluzione delta gen LWL dust}
   &&\Delta_m=K_1 t^{-1}+K_2 t^{\alpha _+|_{w=0}}+K_3 t^{\alpha _-|_{w=0}}-K_4 \frac{\mathcal{C}_0}{a_0^2}
   t^{2-\frac{4 n}{3}}\;,
\end{eqnarray}
where
\begin{eqnarray}
 &&\alpha_{\pm}|_{w=0} = -\frac{1}{2}\pm\frac{\sqrt{(n-1) (n (32 n (8 n-19)+417)-81)}}{6 (n-1)}\;,\\
 &&K_4|_{w=0}= \frac{9 (n (12 n-31)+18)}{8 (4 n-9) \left(12 n^3-19 n^2-3 n+9\right)}\;.
\end{eqnarray}
A graphical representation of the behavior of the exponent of the
modes in \rf{soluzione delta gen LWL dust} as $n$ changes is given
in Figure \ref{plotdust}. This solution has many interesting
features. For $0.33 < n < 0.71$ and $1 < n < 1.32$ \footnote{The
values of $n$ presented here and in the following are the result of
the resolution of algebraic equations of order greater than two.
These values have been necessarily calculated numerically. This
means that the values we will give for the interval $n$ will be
necessarily an approximation.} the modes $t^{\alpha_{\pm}|_{w=0}}$
become oscillatory. However since the real part of the exponents
$\alpha_{\pm}|_{w=0}$ is always negative the oscillation are damped
and bound to become subdominant at late times. The appearence
of this kind of modes is not associated with any peculiar behavior
of the thermodynamic quantities in the background i.e. none of the
energy condition are violated for the values of $n$ which are associated with the
oscillations. The nature of these oscillations is then an higher
order phenomenon. Here we will not undertake a detailed
investigation of the origin of these modes, such a study will be
left for a future work. Also, for most of the values of $n$ the
perturbations grow faster in $R^n$-gravity than in GR. In fact only
for $1.32\leq n < 1.43$ all the modes grow with a rate slower than
$t^{2/3}$.

Probably the most striking feature of the solutions \rf{soluzione
delta gen LWL dust} and  \rf{soluzione delta gen LWL} is that the
long wavelength perturbations grow for every value of $n$, even if
the universe is in a state of accelerated expansion (see Figure
\ref{plotdust}). This is somehow expected from the fact that in
\cite{cdct:dynsys05} the fixed point representing our background is
unstable for every value of the parameters. However, the consequence
of this feature is quite impressive because it implies that in
$R^{n}$ gravity large scale structures can in principle also be
formed in accelerating backgrounds. This is not possible in General
Relativity, where it is well known that as soon as the deceleration
parameter becomes positive the modes of the $\Delta$ solutions (or
density contrast) are both decreasing. The suppression of
perturbations due to the presence of classical forms of Dark Energy
(DE) is one of the most important sources of constraints on the
nature of DE itself. Our example shows that if one considers DE as a
manifestation of the non-Einsteinian nature of the gravitational
interaction on large scales, there is the possibility to have an
accelerated expanding background  that is compatible with the growth
of structures. Of course, in order to better understand this effect,
one should also analyze the evolution of perturbations on small
scales. However this analysis is beyond the scope of this paper and
it is left to left to a future, more detailed investigation.

In the limit $n\rightarrow 1$ two of the modes of \rf{soluzione
delta gen LWL} reproduce the two classical modes $t^{2/3}$ and
$t^{-1}$ typical of GR, but the other two diverge. At first glance
this might be surprising  but it does not represent a real pathology
of the model. In fact equation \rf{EqDelta3Ord} reduces to a first
order differential equation when $n=1$. Therefore in this case the
two modes in the solution can be discarded and GR is recovered.

From the system \rf{sistemLWLBckgr} we can also obtain the solution
for the other scalars:
\begin{eqnarray}\label{soluzione others gen LWL}
   && \mathcal{R}=K_5 t^{\frac{2 n \omega
   }{\omega +1}-3}+K_6 t^{\beta _+}+K_7 t^{\beta _-}-K_8 \frac{\mathcal{C}_0}{a_0^2} t^{-\frac{4 n}{3 (\omega
   +1)}}\\&&\Re= K_9 t^{\frac{2 n \omega
   }{\omega +1}-1}+K_{10} t^{\gamma _+}+K_{11} t^{\gamma _-}-K_{12} \frac{\mathcal{C}_0}{a_0^2}  t^{-\frac{4 n}{3 (\omega
   +1)}-1}
\end{eqnarray}
where
\begin{eqnarray}
  \beta_{\pm} &=& \alpha_{\pm}-2\;,\\
 \gamma_{\pm} &=& \alpha_{\pm}-3
\end{eqnarray}
and the constants $K_5,..K_{12}$ are all functions of $K_1,..K_4$.
These expression are rather complicated and will not be given here. It is
interesting that these quantities have an oscillatory behavior for the
same values of $n$ for which $\Delta_m$ is oscillating. Also for
these quantities the oscillating modes are always decreasing.

Finally it is useful to derive and expression for  the Newtonian
potential $\Phi_N$ given in \rf{eq:BardeenPHI} which for our
background takes the form
\begin{eqnarray}\label{eq:BardeenPHISol}
  \nonumber\Phi_N &=& \frac{4 n a_0^2 K_1 t^{\frac{2 \omega  n}{\omega +1}+\frac{4 n}{3 (\omega
   +1)}-3}}{3 (\omega +1)^2}+\frac{4 n \left(2 n \omega -(\omega +1) \alpha
   _-\right) a_0^2 K_2 t^{\frac{4 n}{3 (\omega +1)}+\beta_-}}{3 (\omega
   +1)^3}\\ &&+\frac{4 n \left(2 n \omega -(\omega +1) \alpha _+\right) a_0^2
   K_3t^{\frac{4 n}{3 (\omega +1)}+\beta _+}}{3 (\omega
   +1)^3}+\frac{9
   (\omega +1)^3-16 n (2 n+3 (n-1) \omega -3) K_4 }{18
   (\omega +1)^3}\mathcal{C}_0\;.
\end{eqnarray}
As in the GR case, this potential has a constant mode and at least
one monotonic mode. The presence of a constant mode it is important
because it is consistent with the standard Sachs-Wolfe plateau in
GR. In addition, the fourth order correction induce oscillations in
two of the modes of \rf{eq:BardeenPHISol} as it is expected from
\rf{eq:BardeenPHI} and the form of the solution for $\Delta_m$. As
for $\Delta_m$ these oscillatory modes always decay. Again, when
$n\rightarrow 1$ the Newtonian potential reduces to the one obtained
in General Relativity.

\begin{figure}
  \includegraphics[width=11cm]{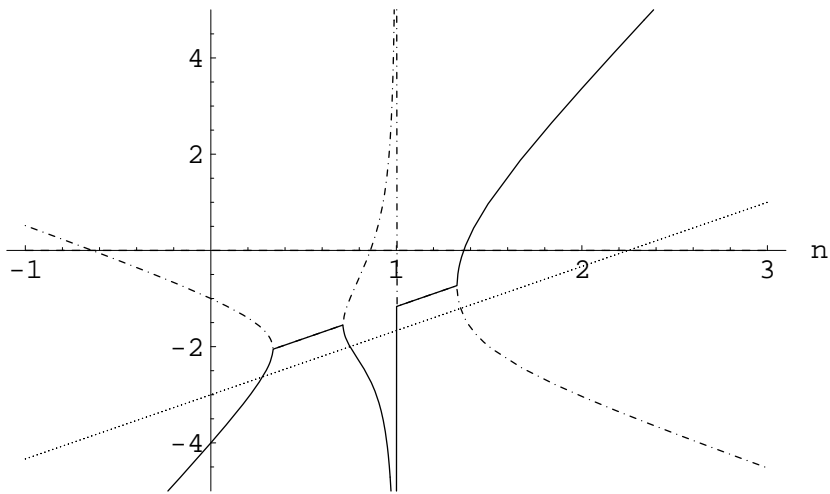}\\
  \caption{Plot of the real part of  exponents of the modes of the
  Newtonian Potential \rf{eq:BardeenPHISolDust} against $n$ in the case of dust ($w=0$).
  The continuous and dot-dashed line represent the modes $t^{\frac{4 n}{3 }+\beta_{\pm}|_{w=0}}$
  respectively (note how they coincide when $\beta_{\pm}$ are complex),
  the dotted line represents the mode $ t^{\frac{4 n}{3 }-3}$.}\label{plotNPdust}
\end{figure}

In the case of dust we have
\begin{eqnarray}\label{eq:BardeenPHISolDust.EPS}
 \Phi_N &=&  \frac{4 n}{3}  a_0^2 K_1 t^{\frac{4 n}{3 }-3}-\frac{4 n
  }{3}\;a_0^2 K_2\;\alpha_-|_{w=0} t^{\frac{4 n}{3 }+\beta_-|_{w=0}}-\frac{4 n
  }{3}\;a_0^2 K_3\;\alpha_+|_{w=0} t^{\frac{4 n}{3 }+\beta_+|_{w=0}}
   +\left[\frac{1}{2}-\frac{8}{9} n (2 n -3) K_4\right] \mathcal{C}_0\;,
\end{eqnarray}
(see Fugure \ref{plotNPdust}). This potential is weaker that the GR
one at early times and became stronger at late times. The non
constant modes are decaying only for $-0.63 < n \leq 0.33$, $
0.71\leq n < 0.86$ and $1.32 \leq n < 1.36$ and their exponent is
grater than the one in GR for any $n$ but $0.28 < n \leq 0.33$.
\section{Conclusion}
In this paper we have analyzed in a rigorous and mathematically well
defined way the evolution of density perturbations of FLRW
backgrounds in fourth order gravity.

Our analysis has been based on two important steps. The first one
follows from the fact that in homogeneous and isotropic spacetimes
the field equations for a generic fourth order gravity theory can be
rewritten in a form that resembles  standard GR plus two effective
fluids. Then, using 1+3 covariant approach, it is possible to derive
the general equations describing the evolution of the cosmological
perturbations of these models in FLRW background. In this paper we
have dealt only with the evolution of the scalar perturbations,  the
evolution  of tensor and vector perturbations will be presented
elsewhere \cite{K1,K2}.

Providing that one has a clear picture in mind of the effective nature of
the fluids involved, the approach above has the advantage of making
the treatment of the perturbations physically clear and
mathematically rigorous. In particular, it allows one
to understand in a natural way that the equations
governing the scalar perturbations in fourth order gravity are
of order four rather than two.

Once the general perturbations equations were derived, we applied
them to the simple $R^{n}$-gravity model. In the long wavelength
limit and using a background solution derived from an earlier
dynamical systems analysis  \cite{cdct:dynsys05},  we were able to
find exact solutions to the perturbations equations. The results we
obtain are particularly interesting. As expected, our background
solution proved to be unstable under scalar perturbations - this
solution always corresponds to a saddle point in the phase space of
the $R^n$ homogeneous and isotropic models \cite{cdct:dynsys05}. In
addition, for specific intervals of values of the parameter $n$ two
of the four modes of the solution can become oscillatory and this
might have  consequences on the scalar perturbation spectrum. The
connection between the spectrum of matter perturbations and the CMB
power spectrum then offers an interesting independent way of testing
these models on cosmological scales

However the most striking property of the evolution of the density
perturbations in this model is that growth is possible even if the
background is accelerating. This means that {\em unlike  all the
other models for dark energy, in $R^{n}$-gravity a decelerated phase
is not necessarily required to form the large scale structure}. Of course
in order to fully support this claim a detailed analysis of the small-scale
perturbations  is necessary in order to understand better, if in this regime
small scale structure formation is also preserved. However, the idea is
very intriguing. A number of important constraints on the nature of
Dark Energy come from the requirement that we have a matter dominated
phase able to support the formation of large-scale structures. What we have discovered
is that  in $R^{n}$-gravity this is not necessarily the case.

Also the Newtonian potential has a number of  interesting features.
First of all it contains a constant mode like in GR. This is an
encouraging result as this is compatible with the Sachs-Wolfe
plateau. On the other hand, the time evolution of the Newtonian
potential is essentially different form GR.  Such differences
suggest that the dynamics of structure formation in $R^{n}$-gravity
might be very different from what we find in GR and it is definitely
worth a more detailed study.

A natural question arises.  How general are these results in
terms of the form of the fourth order Lagrangian? Or in other words
are also other, more popular, fourth order models able to give rise
to the same effects? The question is not easy to answer. From the
form of the decoupled equation for the matter density perturbations,
we can conclude that features like the oscillating modes should be a
common to all fourth order theories. On the other hand proving that
density perturbation grow for all accelerating backgrounds  is a much difficult matter.
One interesting hint comes from the general dynamical system
analysis given in \cite{SanteGenDynSys} in which backgrounds
similar to the one we used for $R^{n}$-gravity are often unstable
for most the values of the parameters of that theory. Finally, from
the point of view of the dynamics of structure formation, it seems to us
reasonable to say that if a model as ``close" to GR as
$R^{n}$-gravity has so many different features, fourth order
gravity models with a Lagrangian very  different from the one in the
Hilbert-Einstein action  will in general have a very
different dynamics. However, as we have seen in $R^{n}$ gravity,
these differences do not necessarily imply a complete incompatibility
with the data coming from the CMB and other observational constraints. However,
much more work will have to be done before we can determine whether alternative gravity
provides a viable alternative to our much cherished theory of General Relativity.

\vfill {\noindent{\bf Acknowledgements:}\\
This work was supported by the National Research Foundation (South
Africa) and the {\it Ministrero deli Affari Esteri- DIG per la
Promozione e Cooperazione Culturale} (Italy) under the joint
Italy/South Africa science and technology agreement. A.T. warmly
thanks the Department of Mathematics and Applied Mathematics UCT and
the University of Cape Town for hospitality and support. The authors
would like to thank Prof Salvatore Capozziello for some preliminary
discussions on this subject and Dr Kishore Ananda for useful
suggestions. Finally P. D. and S. C.  would like to thank the
Institute of Astrophysics (IAP) in Paris for hospitality during the
final stages of this work.
\newpage
\appendix
\section{General propagation and constraint equations}
For a general, imperfect energy-momentum tensor the propagation and constraint equations are:\\

\noindent Expansion propagation (generalized Raychaudhuri equation):
\begin{eqnarray}\label{1+3eqRayHO2}
&&\dot{\Theta}+{\textstyle\frac{1}{3}}\Theta^2+\sigma_{{{a}}{{b}}}
\sigma^{{{a}}{{b}}} -2\omega_{{a}}\omega^{{a}} -\tilde{\nabla}^{{a}}
\dot{u}_{{a}}+ \dot{u}_{{a}} \dot{u}^{{a}}+{\textstyle\frac{1}{2}}(\mu
+ 3 p) \nonumber=0
\end{eqnarray}
Vorticity propagation:
\begin{equation}\label{1+3VorHO2}
\dot{\omega}_{\langle {{a}}\rangle }
+{\textstyle\frac{2}{3}}\Theta\omega_{{a}} +{\textstyle\frac{1}{2}}\curl
\dot{u}_{{a}} -\sigma_{{{a}}{{b}}}\omega^{{b}}=0 \,.
\end{equation}
Shear propagation:
\begin{equation}\label{1+3ShearHO2}
\dot{\sigma}_{\langle {{a}}{{b}} \rangle }
+{\textstyle\frac{2}{3}}\Theta\sigma_{{{a}}{{b}}}
+E_{{{a}}{{b}}}-\D_{\langle {{a}}}\dot{u}_{{{b}}\rangle }
+\sigma_{{c}\langle {{a}}}\sigma_{{{b}}\rangle }{}^{c}+
\omega_{\langle {{a}}}\omega_{{{b}}\rangle} - \dot{u}_{\langle
{{a}}}\dot{u}_{{{b}}\rangle} -\frac{1}{2}\pi_{{{a}}{{b}}}=0
\end{equation}
Gravito-electric propagation:
\begin{eqnarray}\label{1+3GrElHO2}
 && \dot{E}_{\langle {{a}}{{b}} \rangle }
+\Theta E_{{{a}}{{b}}} -\curl H_{{{a}}{{b}}}
+{\textstyle\frac{1}{2}}(\mu+p)\sigma_{{{a}}{{b}}}
 -2\dot{u}^{c}\ep_{{c}{d}({{a}}}H_{{{b}})}{}^{d}
-3\sigma_{{c}\langle {{a}}}E_{{{b}}\rangle }{}^{c} +\omega^{c}
\ep_{{c}{d}({{a}}}E_{{{b}})}{}^{d} \nonumber\\&&~~{}
+{\textstyle\frac{1}{2}}\dot{\pi}_{\langle {{a}}{{b}}\rangle }
+{\textstyle\frac{1}{2}}\D_{\langle {{a}}}q_{{{b}}\rangle }
+{\textstyle\frac{1}{6}}
\Theta\pi_{{{a}}{{b}}}+{\textstyle\frac{1}{2}}\sigma^{c}{}_{\langle
{{a}}}\pi_{{{b}}\rangle {c}} +{\textstyle\frac{1}{2}}
\omega^{c}\ep_{{c}({{a}}}^{d}\pi_{b )d}{}=0
\end{eqnarray}
Gravito-magnetic propagation:
\begin{eqnarray}\label{1+3GrMagHO2}
&&\dot{H}_{\langle {{a}}{{b}} \rangle } +\Theta H_{{{a}}{{b}}}+\curl
E_{{{a}}{{b}}}- 3\sigma_{{c}\langle
{{a}}}H_{{{b}}\rangle}{}^{c}+\omega^{c}\ep_{{c}{d}({{a}}}H_{{{b}})}{}^{d}+2\dot{u}^{c}\ep_{{c}{d}({{a}}}E_{{{b}})}{}^{d}
\nonumber\\&&~~{}={\textstyle\frac{1}{2}}\curl\pi_{{{a}}{{b}}}-{\textstyle\frac{3}{2}}\omega_{\langle
{{a}}}q_{{{b}}\rangle}+{\textstyle\frac{1}{2}}\sigma^{c}{}_{({{a}}}\ep_{{{b}}){c}}^{\;\;\;\;d}q_{d}
\end{eqnarray}
Vorticity constraint:
\begin{equation}\label{1+3VorConstrHO2}
\D^{{a}}\omega_{{a}} -\dot{u}^{{a}}\omega_{{a}} =0\,.
\end{equation}
Shear constraint:
\begin{equation}\label{1+3ShearConstrHO2}
\D^{{b}}\sigma_{{{a}}{{b}}}-\curl\omega_{{a}}
-{\textstyle\frac{2}{3}}\D_{{a}}\Theta +2[\omega,\dot{u}]_{{a}}
-q_{a}=0
\end{equation}
Gravito-magnetic constraint:
\begin{equation}\label{1+3GrMagConstrHO2}
 \curl\sigma_{{{a}}{{b}}}+\D_{\langle {{a}}}\omega_{{{b}}\rangle  }
 -H_{{{a}}{{b}}}+2\dot{u}_{\langle {{a}}}
\omega_{{{b}}\rangle  }=0 \,.
\end{equation}
Gravito-electric divergence:
\begin{eqnarray}\label{1+3GrElConstrHO2}
 && \D^{{b}} E_{{{a}}{{b}}} -{\textstyle\frac{1}{3}}\D_{{a}}\tilde{\mu}^{m}
 -[\sigma,H]_{{a}}+3H_{{{a}}{{b}}}\omega^{{b}}
-{\textstyle\frac{1}{2}}\sigma_{{{a}}}^{{b}}q_{{b}}+{\textstyle\frac{3}{2}}
[\omega,q]_{{a}}+{\textstyle\frac{1}{2}}\D^{{b}}\pi_{{{a}}{{b}}}
-{\textstyle\frac{1}{3}}\D_{{a}}\mu+{\textstyle\frac{1}{3}}\Theta
q_{{a}}=0
\end{eqnarray}
Gravito-magnetic divergence:
\begin{eqnarray}\label{1+3GrMagDivHO2}
&&\D^{{b}} H_{{{a}}{{b}}}-(\mu+p)\omega_{{a}}
+[\sigma,E]_{{a}}-3E_{{{a}}{{b}}}\omega^{{b}}
+{\textstyle\frac{1}{2}}\curl q_{{a}} +{\textstyle\frac{1}{2}}
[\sigma,\pi]_{{a}}-{\textstyle\frac{1}{2}}\pi_{{{a}}{{b}}}
\omega^{{b}}=0
\end{eqnarray}
Here $\omega_{{a}}=\frac{1}{2}\ep_{a}{}^{{b}{c}}\omega_{bc}$ and the
covariant tensor commutator is
\[
[W,Z]_{{a}} =\ep_{{{a}}{c}{d}}W^{c}{}_{e} Z^{{d}{e}}\,.
\]

\section{Covariant formalism versus Bardeen's formalism}
As we have seen the covariant approach is a very useful framework
for studying  perturbations in alternative theories of  gravity.
However, since most work on cosmological perturbations is usually
done using the Bardeen approach \cite{bi:bardeen}, we will give here
a brief summary of how can relate our quantities to the standard
Bardeen potentials. A detailed analysis of the connection between
these formalism is given in \cite{BDE}. Here we limit ourselves to
give the main results  for scalar perturbations.

In Bardeen's approach to perturbations of FLRW spacetime, the metric
$g_{ab}$ is the fundamental object. If $\bar{g}_{ab}$ is the
background metric and $g_{ab}=\bar{g}_{ab} +\delta g_{ab}$ defines
the metric perturbations $\delta g_{ab}$ in these coordinates.

The perturbed metric can be written in the form
\begin{equation}
ds^2= S^2(\eta)\{-(1+2A)d\eta^2 -2 B_\alpha dx^\alpha d\eta
+[(1+2H_L) \gamma_{\alpha\beta}+2 H^T_{\alpha\beta}]dx^\alpha
dx^\beta\}\;, \label{eq:pertmetr}
\end{equation}
where $\eta$ is the conformal time, and the spatial coordinates are
left arbitrary. This spacetime can be foliated in 3-hypersurfaces
$\Sigma$ characterized by constant conformal time $\eta$ and metric
$\gamma\ab$.

The quantities $A$ and $B_\alpha$ are respectively the perturbation
in the lapse function (i.e. the ratio of the proper time distance
and the coordinate time one between two constant time hypersurfaces)
and in the shift vector (i.e. the rate of deviation of a constant
space coordinate line from the normal line to a constant time
hypersurface),  $H_L$ represents the amplitude of perturbation of a
unit spatial volume and $H^T_{\alpha\beta}$ is the amplitude of
anisotropic distortion of each constant time hypersurface
\cite{KodamaSasaki}.

The minimal set of perturbation variables is completed by defining
the fluctuations in the energy density: \ber  \mu=\bar{\mu}
+\delta\mu\;, & \delta\equiv\delta\mu/\bar{\mu}\;,
 \label{eq:dmu}
 \eer
and the fluid velocity: \ber u^a=\bar{u}^a +\delta u^a\;, & \delta
u^\alpha= \bar{u}^0 v^\alpha\;, & \delta u^0=-\bar{u}^0 A \;,
 \label{eq:du}
\eer
 together with the energy flux $q_a$ and the anisotropic
pressure $\pi_{ab}$ which are GI by themselves.

These  quantities are treated as 3-fields propagating on the
background 3-geometry. With suitable choice of boundary conditions
\cite{bi:stewart}, these quantities  can be uniquely (but
non-locally) decomposed into scalars, 3-vectors and 3-tensors: \ber
B_\alpha&=&B_{|\alpha} +B^S_\alpha\;,
 \label{eq:split1} \\
 H_{T\alpha\beta}&=&
\na_{\alpha\beta}H_T +H^S_{T(\alpha|\beta)}
+H^{TT}_{T\alpha\beta}\;,
 \label{eq:split2}
 \eer
 where the slash indicates covariant differentiation with respect to the
the metric $\gamma\ab$ of $\Sigma$. In this way  $\na\ab
f=f_{|\beta\alpha}-{\textstyle\frac{1}{3}} \na^2f$ and
$\na^2f=f^{|\gamma}{}_{|\gamma}$ is the Laplacian.  The superscript
$S$ on a vector means it is solenoidal ($B^{S|\alpha}_{\alpha}=0$),
and $TT$ tensors are transverse
($H^{TT\beta}_{T\alpha}{}_{|\beta}=0$) and trace-free.

On the base of \rf{eq:split1} and \rf{eq:split2}, it is standard  to
define {\em scalar} perturbations as those quantities which are
3-scalars, or are derived from a scalar through linear operations
involving only the metric $\gamma\ab$  and its $|$ derivative.
Quantities derived from similar operations on solenoidal vectors and
on $TT$ tensors are dubbed  {\em vector} and {\em tensor}
perturbations. Scalar perturbations are relevant to matter clumping,
i.e. correspond to density perturbations, while vector and tensor
perturbations correspond to rotational perturbations and
gravitational waves.

Given the homogeneity and isotropy of the background, we can
separate each variable into its time and spatial dependence using
the method of harmonic decomposition. In the Bardeen approach the
standard harmonic decomposition is performed using the
eigenfunctions
 of the Laplace-Beltrami operator on 3-hypersurfaces of constant curvature
$\Sigma$ (i.e. on the homogeneous spatial sections of FLRW
universes). In particular these harmonics are defined by
\begin{eqnarray}
&&\nabla^2 Y^{(k)}=-k^2 Y^{(k)}\;, \label{harmY1} \\
&&\nabla^2 Y^{(k)}_{\alpha}=-k^2Y^{(k)}_{\alpha}\;, \label{harmY2}\\
&& \nabla^2 Y^{(k)}_{\alpha\beta}=-k^2Y^{(k)}_{\alpha\beta}\;,
\label{harmY3}
\end{eqnarray}
where $Y^{(k)},Y^{(k)}_{\alpha},Y^{(k)}_{\alpha\beta}$ are the
scalar, vector and tensor harmonics of order $k$. In this way one
can decompose scalars, vectors and tensors as
\ber f&=&f(\eta)Y\\
B_\alpha&=& B\0(\eta) Y\0_\alpha +B\1(\eta) Y\1_\alpha\;,
 \label{eq:split1harm} \\
 H_{T\alpha\beta}&=&H\0_T(\eta)Y\0_{\alpha\beta}+H\1_T(\eta)
Y\1_{\alpha\beta} +H\2_T(\eta)Y\2_{\alpha\beta}\;.
\label{eq:split2harm} \eer The key property of linear perturbation
theory of FLRW spacetimes, arising from the unicity of the splitting  (\ref{eq:split1}),
(\ref{eq:split2}), is that in any vector and tensor equation the
scalar, vector and tensor parts on each side are separately equal.

All the quantities define above can be decomposed in this way.
However, before proceeding,  one should note that the quantities $A,
B_\alpha, H^L, H^T_{\alpha\beta}, \delta, v^\alpha$ change their
values under a change of correspondence between the perturbed
``world" and the unperturbed background, i.e., under a {\em gauge
transformation}. In order to have a gauge-invariant theory one has
to  look for combinations of these quantities which are gauge
invariant. Bardeen constructed the following GI variables to treat
scalar perturbations \cite{bi:bardeen} (giving only their $k$-space
representation):
 \ber
\Phi_A&=& \left\{A+\case{1}/{k}\left(B\0{}'+\case{S'}/{S} B\0\right)
-\case{1}/{k^2}\left(H\0_T{}''+ \case{S'}/{S}
H\0_T{}'\right)\right\}Y\;,
\label{eq:phia}\\
\Phi_H &=& \left\{H_L +{\textstyle\frac{1}{3}} H_T\0 +\case{S'}/{k
S}\left(B\0-\case{1}/{k} H\0_T{}'\right)\right\}Y\;,
\label{eq:phih} \\
 V_{S}&=&
 \left(v\0-\case{1}/{k} H_T\0{}'\right)Y\0_\alpha\;,\\
\varepsilon_m&=& =\left[\delta(\eta) +3(1+w)\case{S'}/{k
S}(v\0-B\0)\right]Y\;, \label{eq:epsm}
\eer where  the prime denotes
derivative with respect to the conformal time $\eta$. Note that
there is not a preferred choice of GI density perturbation in this
context, as many other GI combinations are possible
\cite{KodamaSasaki}.

The  variables covariantly defined in the main text are, by
themselves, exact quantities (defined in any spacetime) and  are GI
by themselves, therefore, to first order, we can express them   as
linear combinations of Bardeen's GI variables.  In \cite{BDE}
this expansions is given in full generality. Here we will limit
ourselves to a few examples, giving only the scalar contributions
and refer the reader to \cite{BDE} for details.

The scalar part of the shear, electric part of the Weyl tensor,
energy flux and anisotropic pressure are given by

\begin{eqnarray}
\sigma_{\alpha\beta}&=&
- S k V_S\0 Y\0\ab \label{eq:shear} \\
E\ab&=&
{\textstyle\frac{1}{2}}k^2(\Phi_A-\Phi_H)Y\0\ab \label{eq:eweyl}\\
q_{\alpha} &=& S\left[\kappa h
V_S\0+\case{2k}/{S^2}\left(\Phi_H'-\case{S'}/{S}\Phi_A\right)\right]Y\0_\alpha   \label{eq:qal} \\
\pi\ab &=& -\case{k^2}/{S^2}(\Phi_H+\Phi_A)Y\0\ab\;.
\label{eq:piab1}
\end{eqnarray}
while the scalar parts of the energy density, expansion and
3-curvature scalar gradients can be written as
\begin{eqnarray}\dd_\alpha&=&
-ka\,\varepsilon_m(\eta)Y\0_\alpha\;,\\
\label{eq:dgra}
 \zz_\alpha&=&
\left\{-3k\left( \Phi_H'-\case{S'}/{S} \Phi_A\right) +\left[(3K-k^2)
-{\textstyle\frac{3}{2}}
\kappa h S^2\right]V_S\0\right\}Y\0_\alpha\;, \label{eq:zgra} \\
C_\alpha&=& -4 S k(k^2-3K)\left(\Phi_H-\case{S'}/{ka}
V_S\0\right)Y\0_\alpha\label{eq:cgra}\;. \label{eq:cgrah}
\end{eqnarray}
Finally  the key covariant  scalar perturbation variables are given
by
\begin{eqnarray}
\Delta=S\n^a\dd_a&=& -k^2\varepsilon_m(\eta)Y\;,
 \label{eq:deltah} \\
C=S\n^a C_a&=& k^2 S^2 \,R^*(\eta) Y\;. \label{eq:divc}
\end{eqnarray}
The relations above can be used to give an intrinsic physical and
geometrical meaning to Bardeen's variables, and also to recover his
equations. For example, from the \rf{eq:shear} the variables $V_S$
can be recognized as the scalar contribution to the shear.

The variables $\varepsilon_m$ (\ref{eq:epsm}), interpreted by
Bardeen as the usual density perturbation $\delta\mu/\mu$ within the
comoving gauges $v- B=0$, acquire a covariant significance as the
scalar ``potential'' for the fractional density gradient $\dd_a$
(\ref{eq:dgra}). Obviously, this quantity also represents the
potential for the divergence $\Delta$ (\ref{eq:deltah}) of $\dd_a$
(or its harmonic component).

The two independent GI metric potentials $\Phi_A$ and $\Phi_H$ can
be combined in such a way to give  \be
\Phi_\pi={\textstyle\frac{1}{2}} (\Phi_H +\Phi_A)\;,
~~~\Phi_N={\textstyle\frac{1}{2}} (\Phi_A-\Phi_H)\;;
\label{eq:newpot} \ee  the former $\Phi_\pi$ is a stress potential
while the latter $\Phi_N$ plays exactly the role of a Newtonian
gravitational potential. This last interpretation follows directly
through (\ref{eq:eweyl}), where the scalar part of $E\ab$ has
exactly the same form it has in Newtonian theory $E\ab=\na\ab\Phi_N$
\cite{EllisCovariant}, independently of any gauge choice. For a
perfect fluid ($\pi_{ab}=0$) $\Phi_A$ and $\Phi_H$ are proportional
to each other i.e. $\Phi_A=-\Phi_H$,  however in our case, the
``curvature" fluid is imperfect, so this is not the case. In fact,
we can see from equation (\ref{piR}) that the anisotropic pressure
$\pi_{ab}$ is related to the shear of the matter flow and therefore
the scalar potentials $\Phi_A$ and $\Phi_H$ are related to the
Bardeen shear variable $V_S\0$. This relationship can be calculated
explicitly using (\ref{piR}) together with equations
(\ref{eq:shear}), (\ref{eq:piab1}) and the expression for the Ricci
scalar
$R=2\left(\dot{\Theta}+{\textstyle\frac{2}{3}}\Theta^2+{\textstyle\frac{1}{2}}
\tilde{R}\right)$. After a lengthy calculation we obtain
\begin{eqnarray}
\Phi_A+\Phi_H&=&\frac{f''}{f'}\left\{6\Phi_H''+8\frac{S'}{S}\Phi_H'+4(k^2-3K)\Phi_H
+2\frac{S'}{S}(\Phi_A'+3\frac{S'}{S}\Phi_A)\right.\nonumber\\
&+&\left.2k^{-1}\left[(3K-k^2)-\frac{3}{2}ha^2\right]V'_S+
k^{-1}\left[\frac{4}{3}\frac{S'}{S}(k^2-3K)-8S'ah-3S^2h'\right]V_S\right\}\;.
\end{eqnarray}

\section{Covariant identities} \label{CovID}
On a flat Friedmann background, the following covariant linearized
identities hold:
\begin{eqnarray}
\D_{{a}} \dot f &=& (\D_{{a}}
f)^{\displaystyle\cdot}+\frac{1}{3}\Theta\D_{{a}} f- \dot f \dot{u}_{{a}}\,,\\
\D^2(\D_{{a}} f) &=& \D_{{a}}(\D^2 f)-\frac{2K}{S^{2}}\D_{{a}} f+2\dot f\omega_{{a}}\,,\\
\D^2\dot f& =&
(\D^2f)^{\displaystyle\cdot}+\frac{2}{3}\Theta\D^2f-\dot f\D^{{a}} \dot{u}_{{a}}\,,\\
(\D_{{a}} V_{{b}})^{\displaystyle\cdot}&=&\D_{{a}}\dot{V}_{{b}}-\frac{1}{3}\Theta \D_{{a}} V_{{b}}\,,\\
\D_{[{{a}}}\D_{{{b}}]}V_{c}
&=&-\frac{K}{S^{2}}V_{[a}h_{b]c}\;,\\ 
\D^{{b}}\D_{\langle{{a}}}V_{{{b}}\rangle}&=&
{\textstyle\frac{1}{2}}\D^2V_{{a}} +{\textstyle\frac{1}{6}}
\D_{{a}}(\D^{{b}} V_{{b}})+\frac{K}{S^{2}}V_{{a}}\,,\\
(\D_{{a}}W_{{c}{d}})^{\displaystyle\cdot}&=&\D_{{a}}
\dot{W}_{{c}{d}}-\frac{1}{3}\Theta\D_{{a}} W_{{c}{d}}\,,
\end{eqnarray}
where $V_{{a}}=V_{\langle{{a}}\rangle}$ and
$W_{{{a}}{{b}}}=W_{\langle{{a}}{{b}}\rangle}$ are first order
quantities.


\begin{thebibliography}{99}
 \bibitem{revnostra}
 S.~Capozziello, S.~Carloni and A.~Troisi,
 ``Recent Research Developments in Astronomy \&  Astrophysics"-RSP/AA/21 (2003).
  [arXiv:astro-ph/0303041].
\bibitem{Odintsov}
  S.~Nojiri and S.~D.~Odintsov,
  Phys.\ Rev.\  D {\bf 68} (2003) 123512
  [arXiv:hep-th/0307288]
   \bibitem{Carroll}
    S.~M.~Carroll, V.~Duvvuri, M.~Trodden and M.~S.~Turner,
  Phys.\ Rev.\  D {\bf 70}, 043528 (2004).
  [arXiv:astro-ph/0306438].
\bibitem{Capozziello:2005ku} S.~Capozziello, V.~F.~Cardone and A.~Troisi,
 Phys.\ Rev.\ D {\bf 71} (2005) 043503 [arXiv:astro-ph/0501426];
   \bibitem{Capozziello:2006ph} S.~Capozziello, V.~F.~Cardone and A.~Troisi,
   Mon.\ Not.\ Roy.\ Astron.\ Soc.\ {\bf 375} (2007) 1423 [arXiv:astro-ph/0603522])
  \bibitem{Capozziello:2006jj} S.~Capozziello, A.~Stabile and A.~Troisi,
Mod.\ Phys.\ Lett.\ A {\bf 21} (2006) 2291 [arXiv:gr-qc/0603071].
   \bibitem{Capozziello:2006dj} S.~Capozziello, S.~Nojiri, S.~D.~Odintsov and A.~Troisi,
   Phys.\ Lett.\ B {\bf 639} (2006) 135 [arXiv:astro-ph/0604431]
     \bibitem{cdct:dynsys05} S.~Carloni, P.~Dunsby, S.~Capozziello \& A.~Troisi
\cqg 22, 4839 (2005).
  \bibitem{SanteGenDynSys}  S.~Carloni, A.~Troisi and P.~K.~S.~Dunsby,
  arXiv:0706.0452 [gr-qc] submitted to CQG.
  \bibitem{shosho} M.~Abdelwahab, S.~Carloni and P.~K.~S.~Dunsby,
  arXiv:0706.1375 [gr-qc]. Submitted to CQG.
\bibitem{current1} R.~Bean, D.~Bernat, L.~Pogosian, A.~Silvestri and M.~Trodden,
  Phys.\ Rev.\  D {\bf 75}, 064020 (2007).
  [arXiv:astro-ph/0611321].
\bibitem{current2} Y.~S.~Song, W.~Hu and I.~Sawicki,
  Phys.\ Rev.\  D {\bf 75}, 044004 (2007).
\bibitem{current3}  B.~Li and J.~D.~Barrow,
  Phys.\ Rev.\  D {\bf 75}, 084010 (2007).
  [arXiv:gr-qc/0701111].
\bibitem{current4}
  K.~Uddin, J.~E.~Lidsey and R.~Tavakol,
  arXiv:0705.0232 [gr-qc].
\bibitem{bi:bardeen}
J.~ M.~Bardeen,
 {\it Phys. Rev.} D, {\bf22}, 1982 (1980).
\bibitem{BDE}
M.~Bruni,  P.~K.~S.~Dunsby \& G.~F.~R.~Ellis,
Ap. J. {\bf 395} 34 (1992).
\bibitem{EB} G.~F.~R.~Ellis \& M.~Bruni
Phys Rev D {\bf 40} 1804 (1989).
\bibitem{EBH} G.~F.~R.~Ellis, M.~Bruni and J.~Hwang,
  Phys.\ Rev.\  D {\bf 42} (1990) 1035 (1990).
\bibitem{DBE}  P.~K.~S.~Dunsby, M.~Bruni and G.~F.~R.~Ellis,
  Astrophys.\ J.\  {\bf 395}, 54 (1992)
\bibitem{BED} M.~Bruni, G.~F.~R.~Ellis and P.~K.~S.~Dunsby,
 Class.\ Quant.\ Grav.\  {\bf 9}, 921 (1992).
\bibitem{DBBE}   P.~K.~S.~Dunsby, B.~A.~C.~Bassett and G.~F.~R.~Ellis,
  Class.\ Quant.\ Grav.\  {\bf 14}, 1215 (1997)
  [arXiv:gr-qc/9811092].
 \bibitem{GB} I.~Chavel  {\em Riemannian Geometry: A Modern Introduction} (New York: Cambridge University Press, 1994).
\bibitem{Taylor} R.~Maarteens \& D.~R.~Taylor \grg {\bf 26} 599 (1994);
\bibitem{eddington book}A.~S.~ Eddington {\em The mathematical theory
of relativity} (Cambridge:  Cambridge Univ. Press 1952).
\bibitem{HawkEllis}
  S.~W.~Hawking and G.~F.~R.~Ellis,
  ``The Large scale structure of space-time,''
{\it  Cambridge University Press, Cambridge, 1973}.
 \bibitem{EllisCovariant} G.~F.~R.~Ellis \& H van Elst,
``Cosmological Models", Carg\`{e}se Lectures 1998, in Theoretical
and Observational Cosmology, Ed. M Lachièze-Rey, (Dordrecht: Kluwer,
1999), 1. [arXiv:gr-qc/9812046].
\bibitem{ScSante} S.~Carloni, J.~A.~Leach, S.~Capozziello and P.~K.~S.~Dunsby,
  arXiv:gr-qc/0701009.
\bibitem{bi:stewart}
J.~M.~Stewart,  Perturbations of Friedmann\hs Robertson\hs
Walker cosmological models, {\it Class. Quantum Grav.} {\bf 7},
1169 (1990).
\bibitem{Conserved}
  P.~K.~S.~Dunsby and M.~Bruni,
  Int.\ J.\ Mod.\ Phys.\  D {\bf 3} (1994) 443
  [arXiv:gr-qc/9405008].
\bibitem{K1} K.~ Ananda, S.~ Carloni, P.~ K.~ S.~ Dunsby  in preparation.
\bibitem{K2}   P.~ K.~ S.~ Dunsby, S.~Carloni  and K.~Ananda,  in preparation.
\bibitem{KodamaSasaki} H.~Kodama and M.~Sasaki,
  Prog.\ Theor.\ Phys.\ Suppl.\  {\bf 78} (1984) 1.

\bibitem{bi:Bean2007} R.~Bean, D.~Bernat, L.~Pogosian,
A.~Silvestri and M.~Trodden,
  Phys.\ Rev.\  D {\bf 75} (2007) 064020
  [arXiv:astro-ph/0611321].


 \end{thebibliography}
\end{document}